\pdfoutput=1
\documentclass[11pt,twoside,a4paper,cmspaper,final,collab]{cms-tdr}
\def\svnVersion{daf598d}\def\svnDate{2024/07/20}\def\cmsCernNoTag{CERN-EP-2024-026}\def\cmsCernDate{\today}\def\cmsMessage{Published in the European Physical Journal C as \href{http://dx.doi.org/10.1140/epjc/s10052-024-13021-z}{\doi{10.1140/epjc/s10052-024-13021-z}.}}
\begin{document}\cmsNoteHeader{HIG-22-011}

\newcommand\pp{\ensuremath{\Pp\Pp}\xspace}
\newcommand\bb{\ensuremath{\PQb\PAQb}\xspace}
\newcommand\ZZ{\ensuremath{\PZ\PZ}\xspace}
\newcommand\ZH{\ensuremath{\PZ\PH}\xspace}
\newcommand\HH{\ensuremath{\PH\PH}\xspace}
\newcommand\bjet{\PQb jet\xspace}
\newcommand\bjets{\PQb jets\xspace}
\newcommand\btagged{\PQb-tagged\xspace}
\newcommand\btagging{\PQb tagging\xspace}
\newcommand\bbbb{\ensuremath{\PQb\PAQb\PQb\PAQb}\xspace}
\newcommand\mlead{\ensuremath{m_{\mathrm{jj}}^{\text{lead}}}}
\newcommand\msubl{\ensuremath{m_{\mathrm{jj}}^{\text{subl}}}}
\newcommand\mFourJ{\ensuremath{m_{\mathrm{4j}}}}
\newcommand\dRlead{\ensuremath{\Delta \text{R}_{\text{jj}}^{\text{lead}}}}
\newcommand\dRsubl{\ensuremath{\Delta \text{R}_{\text{jj}}^{\text{subl}}}}
\newcommand\XBB{\ensuremath{X_{\mathrm{B}_1 \mathrm{B}_2}}\xspace}
\newcommand\XHH{\ensuremath{X_{\mathrm{\PH\PH}}}\xspace}
\newcommand\XZZ{\ensuremath{X_{\mathrm{\PZ\PZ}}}\xspace}
\newcommand\XZH{\ensuremath{X_{\mathrm{\PZ\PH}}}\xspace}
\newcommand\XHZ{\ensuremath{X_{\mathrm{\PH\PZ}}}\xspace}
\newcommand\dR{\ensuremath{\Delta\text{R}(j,j)}\xspace}
\newcommand\pDfour{\ensuremath{P(\text{D}_{\mathrm{4b}})}}
\newcommand\pTfour{\ensuremath{P(\text{\ttbar}_{\mathrm{4b}})}}
\newcommand\pMfour{\ensuremath{P(\text{M}_{\mathrm{4b}})}}
\newcommand\pDthree{\ensuremath{P(\text{D}_{\mathrm{3b}})}}

\newcommand\pM{\ensuremath{P(\text{M})}}
\ifthenelse{\boolean{cms@external}}{\providecommand{\cmsLeft}{upper\xspace}}{\providecommand{\cmsLeft}{left\xspace}}
\ifthenelse{\boolean{cms@external}}{\providecommand{\cmsRight}{lower\xspace}}{\providecommand{\cmsRight}{right\xspace}}

\cmsNoteHeader{HIG-22-011}
\title{Search for \texorpdfstring{$\PZ\PZ$ and $\PZ\PH$}{ZZ and ZH} production in the \texorpdfstring{$\PQb\PAQb\PQb\PAQb$}{bbbb} final state using proton-proton collisions at \texorpdfstring{$\sqrt{s}=13\TeV$}{sqrt(s) = 13 TeV}}

\date{\today}

\abstract{
A search for $\PZ\PZ$ and $\PZ\PH$ production in the $\PQb\PAQb\PQb\PAQb$ final state is presented, where \PH is the standard model (SM) Higgs boson. The search uses an event sample of proton-proton collisions corresponding to an integrated luminosity of 133\fbinv collected at a center-of-mass energy of 13\TeV with the CMS detector at the CERN LHC. The analysis introduces several novel techniques for deriving and validating a multi-dimensional background model based on control samples in data. A multiclass multivariate classifier customized for the $\PQb\PAQb\PQb\PAQb$ final state is developed to derive the background model and extract the signal. The data are found to be consistent, within uncertainties, with the SM predictions. The observed (expected) upper limits at 95\% confidence level are found to be 3.8 (3.8) and 5.0 (2.9) times the SM prediction for the $\PZ\PZ$ and $\PZ\PH$ production cross sections, respectively. 
}

\hypersetup{
pdfauthor={CMS Collaboration},
pdftitle={Search for ZZ and ZH production in the bbarbbar final state using proton-proton collisions at sqrt(s) = 13 TeV},
pdfsubject={CMS},
pdfkeywords={CMS, four-bjet, ZZ, ZH, 4b, di-Higgs, hemisphere mixing}}

\maketitle

\section{Introduction}

{\tolerance=800
The observation of Higgs boson pair (\HH) production is an important goal of the High-Luminosity LHC (HL-LHC) program~\cite{Cepeda:2019klc}.
This process is sensitive to the self-coupling ($\lambda$) of the Higgs boson, a crucial parameter of the standard model (SM) that has not yet been measured~\cite{DiMicco:2019ngk}.
Estimates of the sensitivity to the \HH processes indicate that the SM cross section is at the edge of what is observable with an integrated luminosity of 3000\fbinv at the HL-LHC~\cite{CMS:2022dwd}.
An observation of \HH production and a measurement of $\lambda$ will require the combination of the $\bb\PGg\PGg$, $\bb\PGt\PGt$, and $\bbbb$ (4\PQb) decay modes~\cite{CMS:2022cpr,ATLAS:2021ifb,CMS:2022hgz,ATLAS:2022xzm,CMS:2020tkr,ATLAS:2023qzf,PhysRevLett.131.041803}.
The 4\PQb channel has the largest \HH decay branching fraction but suffers from a large background composed of jets produced through the strong interaction, referred to as quantum chromodynamic (QCD) multijet events.
This background is challenging to model with simulation; the QCD multijet predictions lack sufficient accuracy and it is not possible to generate sufficiently large samples.
Extracting all available information from the 4\PQb decay mode will thus require the development and validation of a multi-dimensional background model based on control samples in data.
\par}

In previous \HH $\to$ 4\PQb searches~\cite{ATLAS:2023qzf,CMS:2022cpr,ATLAS:2018rnh,CMS:2018qmt,ATLAS:2016paq,ATLAS:2015zug,CMS:2015jal}, the QCD multijet background is determined in a signal-free control region using a variant of the ``ABCD'' or matrix method~\cite{EDLeCren,PhysRevD.44.29,Behnke:1517556}.
The background prediction in the signal region (SR) requires an extrapolation to a different region of phase space.
This extrapolation is a significant source of systematic uncertainty that limits the ultimate sensitivity of the analysis. 
A common approach for assessing this extrapolation uncertainty is to validate the background prediction in a third, statistically independent validation region (VR)~\cite{ATLAS:2023qzf,CMS:2022cpr,ATLAS:2018rnh,CMS:2018qmt,ATLAS:2016paq,ATLAS:2015zug,CMS:2015jal} that is dominated by background.
This strategy can address how accurately the background model extrapolates to a different region of phase space, but does not directly test the extrapolation into the SR.
In addition, it inevitably suffers from a lack of statistical power in the phase space with the highest signal-to-background ratio;
the selection that makes the VR background-dominated depletes the most sensitive phase space of the SR.

This paper introduces a new method to overcome these limitations by validating the background model with data samples obtained using hemisphere mixing~\cite{arXiv:1712.02538,CMS:2018sxu}, referred to as synthetic data samples.
These synthetic data samples allow for the validation of the extrapolation of the background model to the relevant SR and avoid the problem of low statistical power in the most signal-like phase space.
This technique also provides a way to determine the expected variance of the background prediction, resulting from the finite size of the data sample used to fit the model.

The $\ZZ\to 4\PQb$ and $\ZH \to 4\PQb$ processes share the final state and all the experimental challenges of the $\HH \to 4\PQb$ analysis, but have production cross sections that are expected to be 31 (\ZZ)~\cite{CMS:2016ogx} and 8 (\ZH)~\cite{deFlorian:2016spz} times larger than \HH. 
In addition, the \ZZ and \ZH processes are well established experimentally, both having been observed and measured using channels in which the Z decays to leptons~\cite{CMS:2020gtj,ATLAS:2017bcd,CMS:2018nsn,ATLAS:2018kot,CMS:2023vzh}.

This paper presents a search for $\ZZ \to$ 4\PQb and $\ZH \to$ 4\PQb production using 133\fbinv of proton-proton (\pp) collisions at $\sqrt{s} = 13\TeV$, collected by the CMS experiment at the LHC.
The analysis introduces several new techniques for deriving and validating the background model.
A multiclass multivariate classifier, which uses convolutional layers to solve the combinatoric jet-pairing problem, has been designed with an architecture customized to the 4\PQb final state.
The classifier is used both for signal-versus-background discrimination as well as for the derivation and validation of the background model.
While these techniques are developed and demonstrated in the \ZZ and \ZH $\to$ 4\PQb searches, they are directly applicable to the $\HH \to 4\PQb$ analysis and the measurement of $\lambda$.

The remainder of the paper is organized as follows.
The CMS detector and the reconstruction and identification of physics objects used in this analysis are described in Section~\ref{sec:CMS}.
Section~\ref{sec:data} discusses the data and the simulated events used.
Details of the event selection are presented in Section~\ref{sec:selection}.
Section~\ref{sec:classifier} describes the architecture of the multivariate classifier used throughout the analysis.
The background modeling method is described in Section~\ref{sec:background} and its validation in Section~\ref{sec:closureTest}.
The construction of the synthetic data samples are described in Section~\ref{sec:mixing}, and the evaluation of the background uncertainties in Section~\ref{sec:bkgSys}.
Other sources of systematic uncertainty are detailed in Section~\ref{sec:systematics}.
Finally, the results are reported in Section~\ref{sec:results} and a summary is provided in Section~\ref{sec:summary}.
The tabulated results are provided in a HEPData record~\cite{hepdata}.

\section{The CMS detector}\label{sec:CMS}

The CMS apparatus~\cite{CMS:2008xjf,CMS:2309.05466} is a multipurpose, nearly hermetic detector, designed to trigger on~\cite{Sirunyan:2020zal,Khachatryan:2016bia} and identify electrons, muons, photons, and (charged and neutral) hadrons~\cite{CMS:2020uim,CMS:2018rym,CMS:2014pgm}.
The central feature of the CMS detector is a superconducting solenoid, providing a magnetic field of 3.8\unit{T}.
Within the solenoid volume are a silicon pixel and strip tracker, a lead tungstate crystal electromagnetic calorimeter, and a brass and scintillator hadron calorimeter, each composed of a barrel and two endcap sections.
Forward calorimeters extend the pseudorapidity ($\eta$) coverage provided by the barrel and endcap detectors.
Muons are detected in gas-ionization chambers embedded in the steel flux-return yoke outside the solenoid.
A more detailed description of the CMS detector, together with a definition of the coordinate system used and the relevant kinematic variables, can be found in Ref.~\cite{CMS:2008xjf}.

Events of interest are selected using a two-tiered trigger system.
The first level (L1), composed of custom hardware processors, uses information from the calorimeters and muon detectors to select events at a rate of around 100\unit{kHz} within a fixed latency of 4\mus~\cite{Sirunyan:2020zal}.
The second level, known as the high-level trigger (HLT), consists of a farm of processors running a version of the full event reconstruction software optimized for fast processing, and reduces the event rate to around 1\unit{kHz} before data storage~\cite{Khachatryan:2016bia}.

The event reconstruction is based on the particle-flow (PF) algorithm~\cite{CMS:2017yfk}, which aims to reconstruct and identify each individual particle (PF candidate) in an event with an optimized combination of information from the various elements of the CMS detector.
The PF candidates are classified as electrons, muons, photons, and charged or neutral hadrons.
The primary vertex is taken to be the vertex corresponding to the hardest scattering in the event, evaluated using tracking information alone, as described in Section 9.4.1 of Ref.~\cite{CMS-TDR-15-02}.

Jets are reconstructed from PF candidates, clustered using the anti-\kt algorithm~\cite{Cacciari:2008gp,Cacciari:2011ma} with a distance parameter of 0.4.
The jet momentum is determined as the vectorial sum of all particle momenta in the jet, and is found from simulation to be, on average, within 5--10\% of the true momentum over the whole \pt\ spectrum and detector acceptance. 
Additional collisions within the same or nearby bunch crossings (pileup) can give rise to jets not coming from the hard-scattering process or contribute additional tracks and calorimetric energy depositions, increasing the apparent jet momentum.
To mitigate this effect, tracks identified as originating from pileup vertices are discarded and an offset correction is applied to account for remaining contributions~\cite{Sirunyan:2020foa}.
Jet energy corrections are derived from simulation studies so that the average measured energy of jets becomes identical to that of particle-level jets.
In situ measurements of the momentum balance in dijets, $\gamma+\mathrm{jets}$, $\PZ+\mathrm{jets}$, and multijet events are used to determine any residual differences between the jet energy scale in data and in simulation, and appropriate corrections are applied~\cite{Khachatryan:2016kdb,CMS-DP-2021-033}.
Jets originating from \PQb{} quarks are identified using the \textsc{DeepJet} algorithm~\cite{Bols:2020bkb}, a deep neural network combining secondary vertex properties, track-based variables, and PF jet constituents (neutral and charged particle candidates).
The efficiency of \PQb (light flavor and gluon) jets  to pass the \PQb-tagging requirement used in this analysis is data-set dependent and varies within the 50--60\% (0.05--0.5\%) range.

\section{Data and simulation}\label{sec:data}

This analysis is performed on data collected during three years of data taking at $\sqrt{s}$ = 13\TeV.
The combined data set corresponds to an integrated luminosity of 133\fbinv collected during 2016--2018~\cite{CMS-LUM-17-003,CMS-PAS-LUM-17-004,CMS-PAS-LUM-18-002}.

Events are selected at L1 using triggers requiring the presence of at least four jets in the tracker acceptance ($\abs{\eta} < 2.5$) and large \HT, defined as the scalar sum of the  \pt of the reconstructed jets in the event.
During the 2016 data taking, events are required to have either $\HT > 280\GeV$ or at least four jets with  $\pt > 50\GeV$.
In the 2017 data set, events are required to have $\HT > 280\GeV$ and the four leading jets are required to pass staggered \pt thresholds of 70, 55, 40, and 35\GeV.
In the 2018 data set, the \HT requirement was raised to 320\GeV and the lowest jet \pt threshold was raised to 40\GeV.

Events are selected in the HLT using a combination of triggers requiring the presence of jets coming from the hadronization of \PQb quarks (\PQb jets).
Events are required to have at least four jets, at least three of which are identified as arising from a bottom quark ($\PQb$ tagged).
In the 2016 data set, events are required to have either at least four jets with transverse momentum $\pt > 45\GeV$, or two or more jets with $\pt > 90\GeV$ and two or more jets with $\pt > 30\GeV$.
In the 2017 data set, an \HT requirement of 300\GeV was added to match the threshold at L1, and the four highest-\pt jets were required to pass staggered \pt thresholds of 75, 60, 45, and 40\GeV.
The \HT threshold was raised to 330\GeV in 2018.
The \btagging was performed in HLT using the \textsc{CSV} algorithm~\cite{CMS:2017wtu} in 2016--2017, and with the \textsc{DeepCSV} algorithm~\cite{Bols:2020bkb} in 2018.
Following the selection described in Section~\ref{sec:selection}, this combination of triggers has an approximate efficiency of  20\% for simulated signals with di-boson invariant mass near 200\GeV, rising to 90\% efficiency for masses greater than 600\GeV.

Simulated events  are used to model \ZZ, \ZH, and \HH events and the background from top quark pair (\ttbar) production. 
The \ttbar process is generated at next-to-leading-order (NLO) accuracy in QCD~\cite{Frixione:2007nw}  with \POWHEG v2.0~\cite{Frixione_2007,Campbell_2015}.
The dominant background from QCD multijet events is modeled using control samples in data.

Events from \ZZ production are generated with \MGvATNLO  v2.4.2~\cite{Alwall_2014} at NLO QCD with the FxFx merging scheme~\cite{Frederix:2012ps} and include up to two additional partons.
The SM prediction for the total \ZZ production cross section in \pp collisions at $\sqrt{s} = 13\TeV$ is $15.0^{+0.7}_{-0.6}$\unit{pb}, taken from Ref.~\cite{CMS:2016ogx}.

The quark-induced \ZH signal process is generated at NLO accuracy~\cite{Mimasu:2015nqa} using the \POWHEG v2 event generator extended with the MiNLO procedure~\cite{Hamilton:2012np,Luisoni:2013cuh}, while the gluon-induced \ZH process is simulated at LO accuracy with \POWHEG v2.
The SM prediction for the total \ZH production cross section, computed at next-to-next-to-leading-order accuracy in QCD, is $0.88 \pm 0.03$\unit{pb} for $m_{\mathrm{H}} = 125\GeV$~\cite{deFlorian:2016spz}.
The SM branching fractions of 58.2 and 15.1\% are taken for H $\to$ bb and {Z $\to$ bb} decays, respectively~\cite{deFlorian:2016spz,DJOUADI2019214,PDG2022}, again assuming $m_{\mathrm{H}} = 125\GeV$.

Events from \HH production used to train the classifier are simulated at NLO accuracy ~\cite{Heinrich:2017kxx} with \POWHEG v2.
The dominant SM \HH production mode in \pp collisions at $\sqrt{s} = 13\TeV$ is through the gluon-fusion mechanism; the predicted cross section, computed at next-to-next-to-leading-order accuracy, is $31.1^{+2.1}_{-7.2}$\unit{fb}~\cite{Dawson:1998py,Borowka:2016ehy,Baglio:2018lrj,deFlorian:2013jea,Shao:2013bz,deFlorian:2015moa,Grazzini:2018bsd,Baglio:2020wgt}.

{\tolerance=1600
For both signal and background events, multiple-parton interactions, parton shower, and hadronization effects are simulated with \PYTHIA v8.226 for 2016 and \PYTHIA v8.230 for 2017--2018~\cite{Sj_strand_2015}.
For 2016, the CP5 tune~\cite{Sirunyan_2020} is used for the \ttbar\ samples and the CUETP8M1 tune~\cite{Khachatryan_2016} is used for all the others; for 2017--2018, the CP5 tune is used throughout all samples.
The NNPDF~3.0~\cite{NNPDF:2014otw} (NNPDF~3.1~\cite{NNPDF_3_1}) parton distribution functions (PDFs) are used to simulate the samples corresponding to the 2016 (2017--2018) data sets.
For all simulated samples, the CMS detector response is modeled with \GEANTfour~\cite{AGOSTINELLI2003250}.
Pileup collisions are simulated and added to the hard-scattering process for all samples.
The simulated events are weighted to match the distribution of reconstructed primary vertices observed in data.
\par}

\section{Event selection}\label{sec:selection}

Selected events must have at least four jets with $\pt > 40$\GeV and $\abs{\eta} < 2.4$ that are \btagged by the \textsc{DeepJet} algorithm.
Events passing this selection are referred to as the four-tag sample.
The four jets with the highest \btagging\ score are paired to form \PZ or \PH candidates (``boson-candidate jets'' in what follows).
A dedicated correction to the \PQb jet energy scale, based on a regression technique that takes properties of the jets into account,  is applied to the boson-candidate jets.
This improves the determination of the jet momentum by up to 15\%~\cite{CMS:2017odg}.

Initially, all three possible jet pairings are considered.
To reduce the combinatorial background, pairs of jets with an invariant mass roughly consistent with the \PZ and \PH masses are retained.
Jet pairings must satisfy
\begin{equation}\begin{aligned}
        \label{eqn:SB}
      52 &< \mlead < 180\GeV \quad\text{and} \\
      50 &<  \msubl < 173\GeV,
\end{aligned}\end{equation}
where \mlead\ and \msubl\ are the invariant masses of the leading and subleading boson candidates, respectively.
The leading boson candidate is defined as the one with the highest scalar sum of jet \pt.
From simulation it is found that this sorting tends to bias the leading dijet mass distribution upwards by 2\% and the subleading --- downwards by 2\%.
This is taken into account when defining the dijet mass regions used in the analysis.

{\tolerance=800
The angle between the decay products of the bosons in the laboratory frame provides another handle to reduce the background.
This angle depends on the Lorentz boost of the bosons and, thus, on the four-jet invariant mass \mFourJ.
The pairings of jets associated with the boson candidates satisfy the following requirements
\ifthenelse{\boolean{cms@external}}{
\begin{equation}\begin{aligned}
     \label{eqn:MDRs}       
       \frac{360\GeV}{\mFourJ} -  0.5  < & \dRlead < \max\bigg[1.5, \frac{650\GeV}{\mFourJ} +  0.5 \bigg]\\
       \noalign{\hbox{and}\kern-\baselineskip}\\
       \frac{235\GeV}{\mFourJ} < & \dRsubl < \max\bigg[1.5, \frac{650\GeV}{\mFourJ} +  0.7 \bigg],
          \end{aligned}\end{equation}
}{
\begin{equation}\begin{aligned}
    \label{eqn:MDRs}       
      \frac{360\GeV}{\mFourJ} -  0.5  < & \dRlead < \max\bigg[1.5, \frac{650\GeV}{\mFourJ} +  0.5 \bigg]\quad\text{and}\\
      \frac{235\GeV}{\mFourJ} < & \dRsubl < \max\bigg[1.5, \frac{650\GeV}{\mFourJ} +  0.7 \bigg],
         \end{aligned}\end{equation} 
}
where \dRlead\ and \dRsubl\ are the angular separations between the jets in the leading and subleading boson candidates, respectively.
The angular separation is defined as $\Delta R = \sqrt{\smash[b]{(\Delta \eta)^2 + (\Delta \phi)^2}}$, where $\Delta \phi$ ($\Delta \eta$) is the difference in azimuthal angle (pseudorapidity) between the two boson candidates. 
These requirements reject jet pairings that are inconsistent with a \PZ or \PH decay.
This selection is based on that of the previous ATLAS measurement in Ref.~\cite{ATLAS:2018rnh}, loosened to increase the signal acceptance.
\par}

Events in which all pairings fail the criteria in Eq. (\ref{eqn:MDRs}) are retained to increase the size of the data set used to train various classifiers in the analysis.
If a pairing in these events satisfies one of the aforementioned requirements, it is chosen.
When multiple pairings pass the same number of requirements, one is chosen randomly to give each event a location in the \mlead-\msubl\ plane without biasing the background distribution.

\begin{figure}[hbt!]\centering
\includegraphics[width=0.49\textwidth]{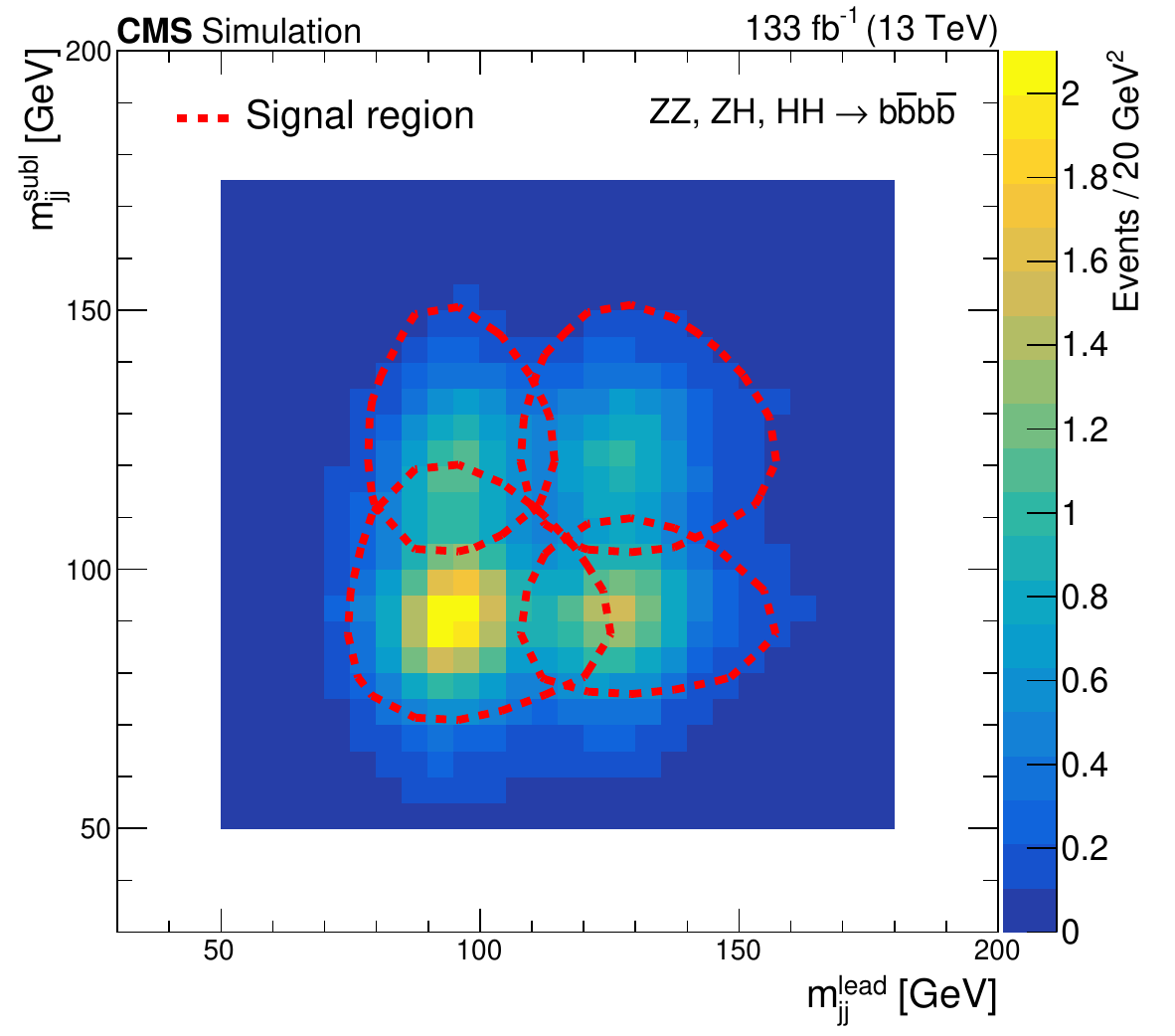}
\includegraphics[width=0.49\textwidth]{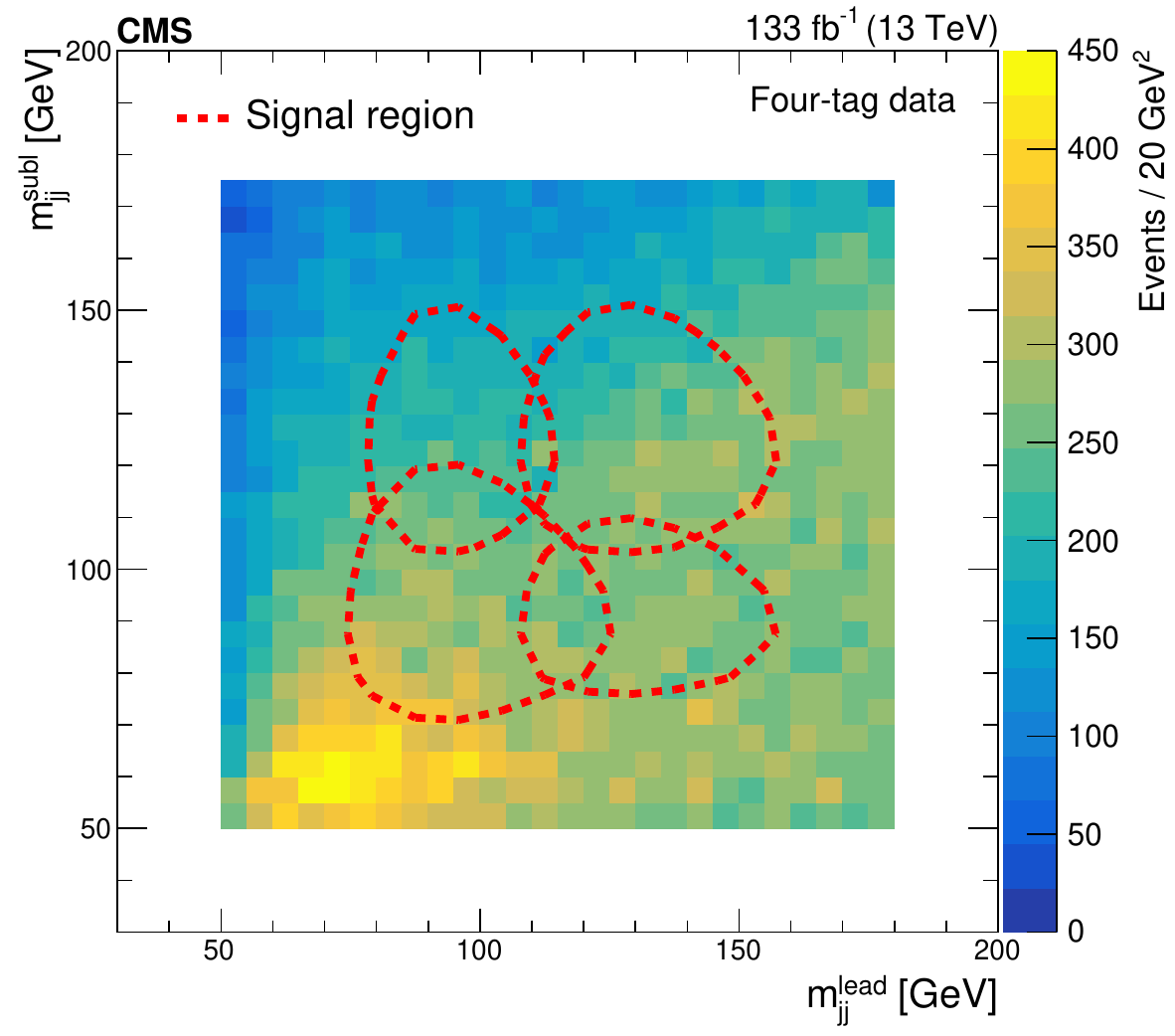}
  \caption{Signal yield from simulation  (\cmsLeft) and from four-tag events in data (\cmsRight), as a function of \mlead and \msubl.
           The color scale to the right of each plot gives the range of values. The signal region is defined by the union of the regions enclosed by the dashed red contours.}
  \label{fig:SBandSR}
\end{figure}

The SR is defined using four overlapping regions in the dijet mass plane targeting \ZZ, \ZH, and \HH decays.
These regions are defined using variables \XZZ, \XZH, \XHZ, and \XHH, defined as
\begin{equation}
  \label{eqn:XBB}
  \XBB = \sqrt{ \Bigg(\frac{\mlead - m_{\mathrm{B_1}}}{\sigma_{\mlead}}\Bigg)^2 + \Bigg(\frac{\msubl - m_{\mathrm{B_2}}}{\sigma_{\msubl}}\Bigg)^2}, 
\end{equation}
where $m_{\mathrm{B_1}}$ and $m_{\mathrm{B_2}}$ correspond to the nominal \PZ\ or \PH\ masses, corrected for the bias mentioned above, depending on the signal targeted.
The denominators in each term is the approximate mass resolution, estimated from simulation to be 10\% of the reconstructed boson mass.
The SR is defined as the union of the requirements: $\XZZ<2.6$, $\XZH<1.9$, $\XHZ<1.9$, and $\XHH<1.9$.
Figure~\ref{fig:SBandSR} shows signal yield from simulation normalized to the expected yield (\cmsLeft) and the four-tag events from data (\cmsRight), as a function of \mlead\ and \msubl.
The four regions used to define the SRs are shown by the red dashed contours.

\begin{figure}[hbt!]\centering
  \includegraphics[width=0.49\textwidth]{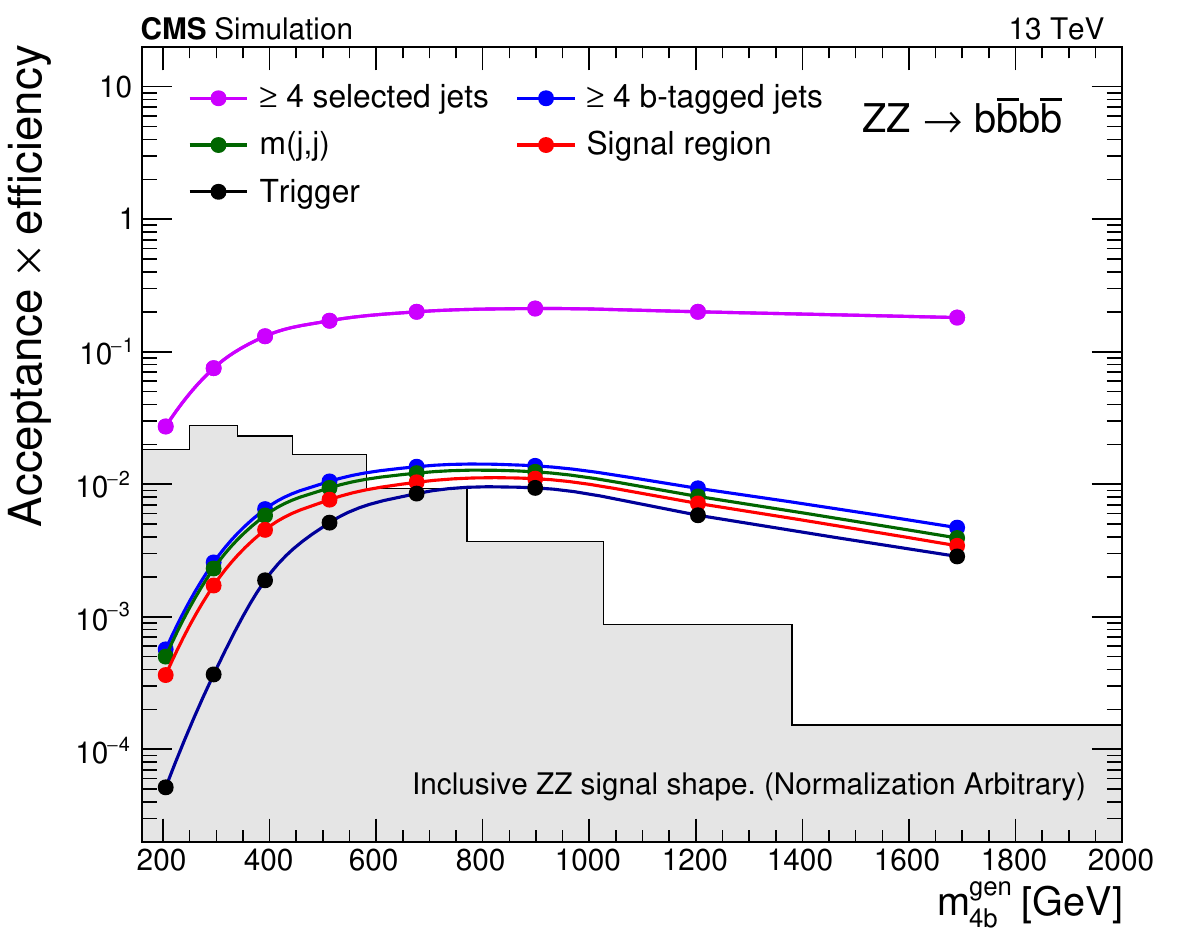}\hfill
  \includegraphics[width=0.49\textwidth]{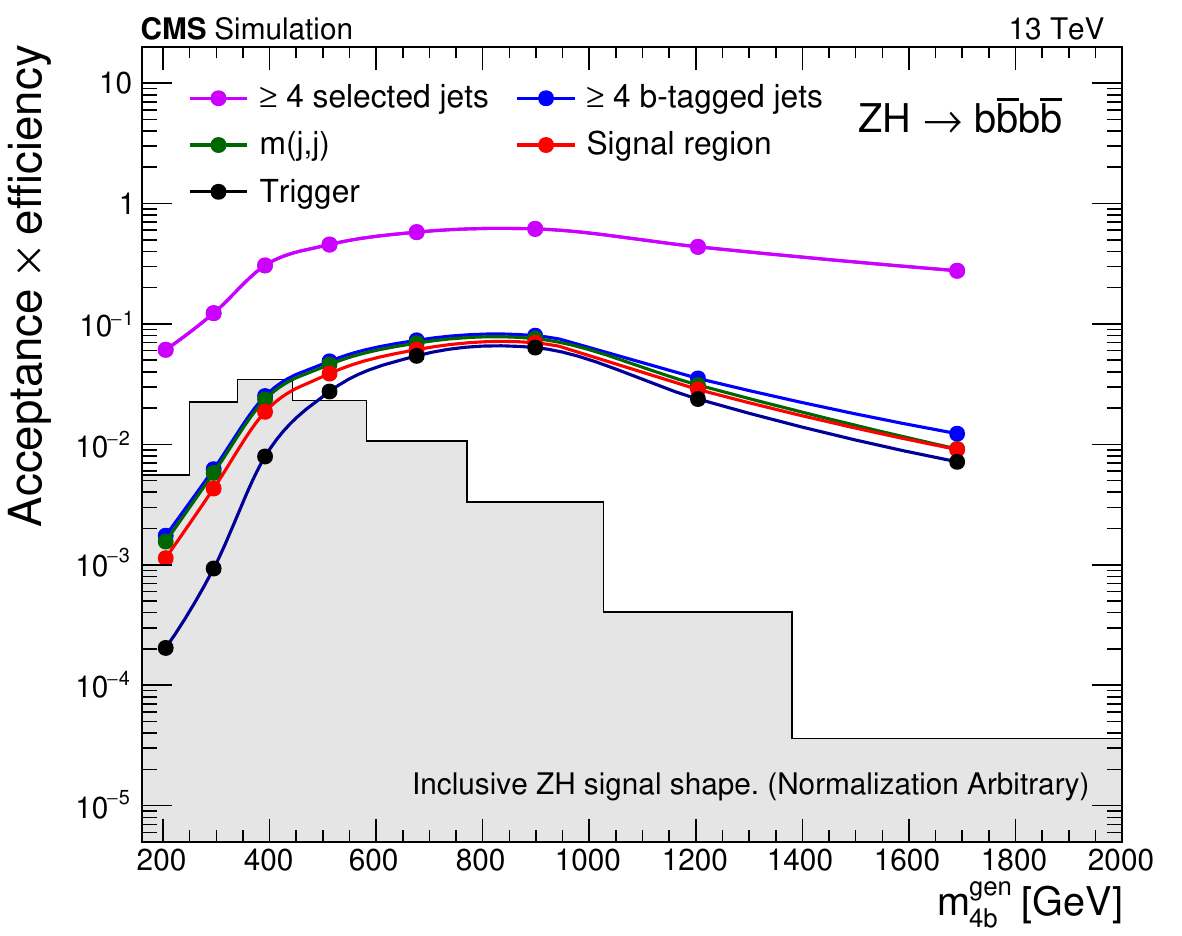}
  \caption{Event selection acceptance times efficiency as a function of the generated four-body mass $m_{4\PQb}^{\text{gen}}$ for the \ZZ\ (\cmsLeft) and \ZH (\cmsRight) signals.
    The plots show the cumulative efficiency with respect to the inclusive sample.
    The expected $m_{4\PQb}^{\text{gen}}$ distributions of the inclusive \ZZ\ and \ZH\ events are shown by the gray-shaded areas with arbitrary normalization.
  }
  \label{fig:accxEff}
\end{figure}

The fractions of \ZZ\ and \ZH\ signal events within the detector acceptance from simulation, multiplied by the efficiency of each selection step are shown in the \cmsLeft and \cmsRight plots of Fig.~\ref{fig:accxEff}, respectively, as a function of the true four-body invariant mass, $m_{4\PQb}^\text{gen}$.
The cumulative effect of each selection requirement is shown for the \ZZ and \ZH channels.
The product of the acceptance and efficiency is limited at low mass by the \HT and jet \pt\ requirements in the trigger.
At high mass, the acceptance for resolving four jets drops and \btagging\ efficiency decreases for high \pt\ jets. 
The trigger requirement significantly limits the efficiency at lower masses.
The larger cross section of the \ZZ process compensates for the lower acceptance, leading to similar expected event yields.

The output from a multivariate classifier, described in Section~\ref{sec:classifier}, is used as the final discriminant between the signal and background.
The classifier, referred to as the signal-versus-background (SvB) classifier,  is trained to output the probability that an event is from one of five classes: multijet, \ttbar, \ZZ, \ZH, and \HH.
The probability of a signal event, defined as the sum of the probabilities of the event being from the \ZZ, \ZH, or \HH sample, is used in the final signal extraction. 

Orthogonal \ZZ\ and \ZH\ SRs are defined according to which of the signal probabilities is the largest.
A combined fit is performed in the \ZZ and \ZH regions to events with a signal probability larger than 1\%, to extract the \ZZ and \ZH fitted signal strengths.
The \HH results are not reported here as a dedicated measurement has been published in Ref.~\cite{CMS:2022cpr}.

\section{Hierarchical combinatorial residual network}\label{sec:classifier}

A multivariate classifier is used both for signal-versus-background discrimination as well as for the derivation and validation of the background model.
An architecture, referred to as the hierarchical combinatorial residual (HCR) network, is specifically developed for the four-jet diboson topology.
The HCR consists of a series of convolutional neural networks, each employing phase-symmetric convolutional filters~\cite{DBLP:journals/corr/ShangSAL16} and residual learning~\cite{he2015deep}, to process kinematic information from the boson-candidate jets; information from additional jets in the event is included using an attention block~\cite{DBLP:journals/corr/VaswaniSPUJGKP17}.
This network architecture provides a clean solution to the combinatorial jet-pairing problem, allowing the same set of weights to be optimized for all pairings. 
The convolutional layers are arranged hierarchically, first processing a jet image to form a dijet image, then processing the dijet image to form a quadjet image. 

\begin{figure*}[hbtp]\centering
  \includegraphics[width=\textwidth]{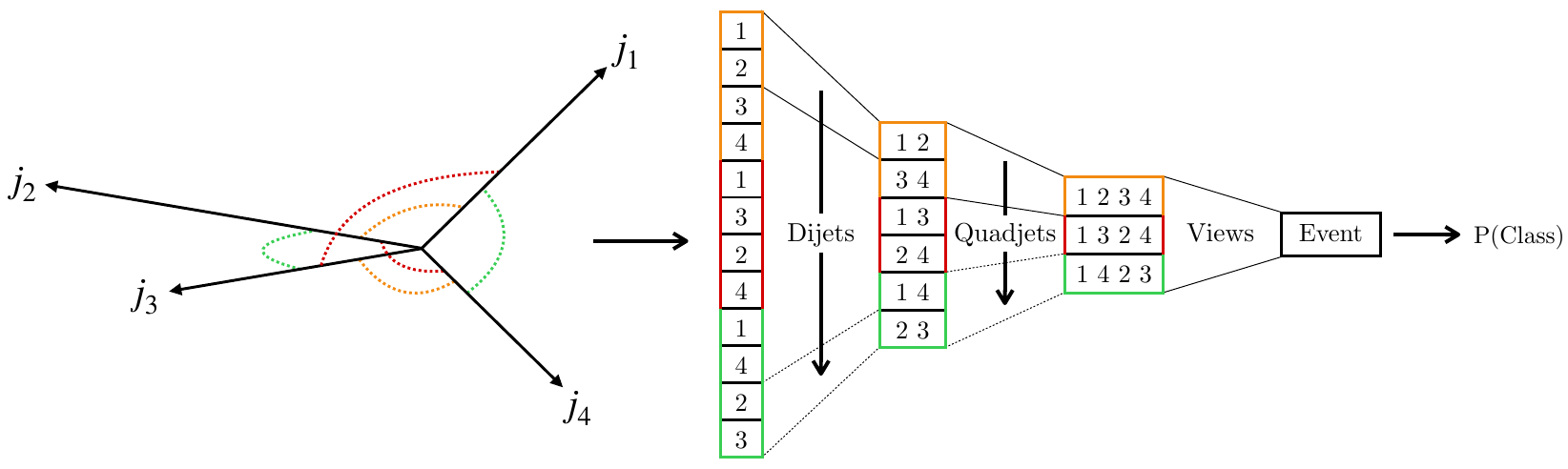}
  \caption{A high-level sketch of the HCR classifier architecture.
           Boson-candidate jets are shown on the left with the three possible jet pairings.
           The HCR architecture is shown on the right.
           The boxes represent pixels, with the labels indicating which jet, dijet, or quadjet the pixel refers to.
           The different jet pairings on the left are each represented within the network, as indicated by the color coding.
           The output P(class) corresponds to the the probability that an event belongs to the corresponding class used in training.
           }
  \label{fig:sketchHCR}
\end{figure*}

Figure~\ref{fig:sketchHCR} shows a high-level sketch of the HCR architecture.
The initial jet image is formed from pixels representing each of the boson-candidate jets, using the jet $\pt,\,\eta,\,\phi,$ and mass values as features.
Three copies of the jet pixels are arranged to form a one-dimensional image such that pairs of adjacent pixels represent the three possible jet pairings.
Adjacent pairs of jet pixels are convolved to form a dijet image using a single set of weights.
The second layer processes the six dijet pixels to form a three-pixel quadjet image.
These three pixels are then combined to produce a single event-level pixel.
Each convolution projects the input image into an eight-dimensional space; the dimension of this embedded space is a hyperparameter that controls the size of the network.
A final output layer projects these features into a $N_{c}$-dimensional space, where $N_{c}$ is the number of input classes used in training.
These outputs are converted into the probabilities that a given event belongs to the corresponding class used in training by applying softmax function. 

{\tolerance=1600
Classifiers with the HCR architecture are used throughout the analysis.
The SvB classifier employs the HCR architecture to construct the variable used in the final signal extraction.
The HCR classifiers are also used to define the background model (Section~\ref{sec:background}) and in the construction of the synthetic data sets (Section~\ref{sec:mixing}).
\par}

\section{Background model}\label{sec:background}

From simulation, it is found that 95\% of the expected background consists of multijet events, which are modeled using data. 
The remaining 5\% are \ttbar\ events, which are modeled with simulation.
Backgrounds from other sources, including processes involving single- and double-\PH production, are found to have a negligible contribution.

The QCD multijet background is modeled with an independent data set selected using the same trigger and selection requirements as used in the SR, except for the \PQb tagging requirement: at least four jets with $\pt > 40\GeV$ are required,  exactly three of which are required to be \PQb tagged.
To increase the size of this three-tag control sample, the \PQb tagging requirement is loosened such that the efficiency to correctly identify a \PQb jet is 70--80\% depending on the data set.
The size of three-tag sample is about 17 times larger than that of the four-tag sample.
The three-tag sample consists of 90\% multijet events and 10\% \ttbar\ events.
Simulations indicate that the three-tag sample has a negligible contribution of signal events.

A product of two weights is applied to the three-tag events to model the multijet background in the four-tag sample.
The first weight, referred to as the  jet combinatorial model (JCM) weight, accounts for additional jet activity in the four-tag sample.
The second weight, referred to as the four-vs-three (FvT) weight,  corrects for kinematic differences between the three- and four-tag samples.

The weights are derived using a signal-depleted sideband (SB) region of the $\mlead-\msubl$\ plane, shown in Fig.~\ref{fig:SBandSR}.
The SB is defined as the region inside the mass window defined by Eq.~(\ref{eqn:SB}) but outside the SR.
The boundaries of the SB region are chosen to provide sufficient statistical precision, while ensuring that the kinematic properties of events in the SB are representative of those in the SR.

\subsection{Jet combinatorial model}\label{sec:JCM}
The four-tag sample has a larger jet multiplicity than the three-tag sample, since the requirement of exactly three \btagged jets in the three-tag sample biases the jet multiplicity.
This effect is modeled using an extension of the combinatorial technique introduced in Ref.~\cite{ATLAS:2018rnh}.
For each three-tag event, all possible combinations of jets not satisfying the looser \PQb tagging requirement, referred to as anti-b-tagged jets, are considered.
At least one anti-\Pb-tagged jet is treated as a \bjet.
The anti-\Pb-tagged jets that are treated as \bjets are referred to as pseudo-tagged jets.
A constant per-jet transfer factor $f$ is assigned to each pseudo-tagged jet and a factor $(1-f)$ to the remaining anti-\Pb-tagged jets.
Correlations in the transfer factor among jets arise because \bjets are produced in pairs.
These correlations are modeled with a pair-enhancement term $e$, that is included in combinations with an even number of tagged plus pseudo-tagged jets.
This enhancement term is expected to be suppressed at higher jet multiplicities since the probability of observing an odd number of \btagged jets increases because of mistagging and jets falling outside of the detector acceptance.
The pair-enhancement factor is thus divided by $n^d$, where $n$ is the number of anti-b-tagged jets and $d$ is a free parameter that controls the suppression of $e$ with jet multiplicity.  
Finally, an overall normalization $t$ is included to account for the looser \PQb tagging requirement in the three-tag sample.
The per-event JCM weight is thus computed as:
\ifthenelse{\boolean{cms@external}}{
\begin{multline}
  \label{eqn:JCM}
  w_{\text{JCM}} \\
  = \begin{cases}
  t\sum_{i=1}^{n} \binom{n}{i}  f^{i}(1-f)^{n-i} (1+ \re/n^d) & (3+i)\quad\text{even} \\ 
  t\sum_{i=1}^{n} \binom{n}{i}  f^{i}(1-f)^{n-i}            & (3+i)\quad\text{odd}, 
\end{cases}
\end{multline}
}{
\begin{equation}
    \label{eqn:JCM}
    w_{\text{JCM}}
    = \begin{cases}
    t\sum_{i=1}^{n} \binom{n}{i}  f^{i}(1-f)^{n-i} (1+ \re/n^d) & (3+i)\quad\text{even} \\ 
    t\sum_{i=1}^{n} \binom{n}{i}  f^{i}(1-f)^{n-i}            & (3+i)\quad\text{odd}, 
  \end{cases}
  \end{equation} 
}
where $i$ is the number of pseudo-tagged jets and $\binom{n}{i}$ is the binomial coefficient.

\begin{figure}[hbt!]\centering
  \includegraphics[width=0.49\textwidth]{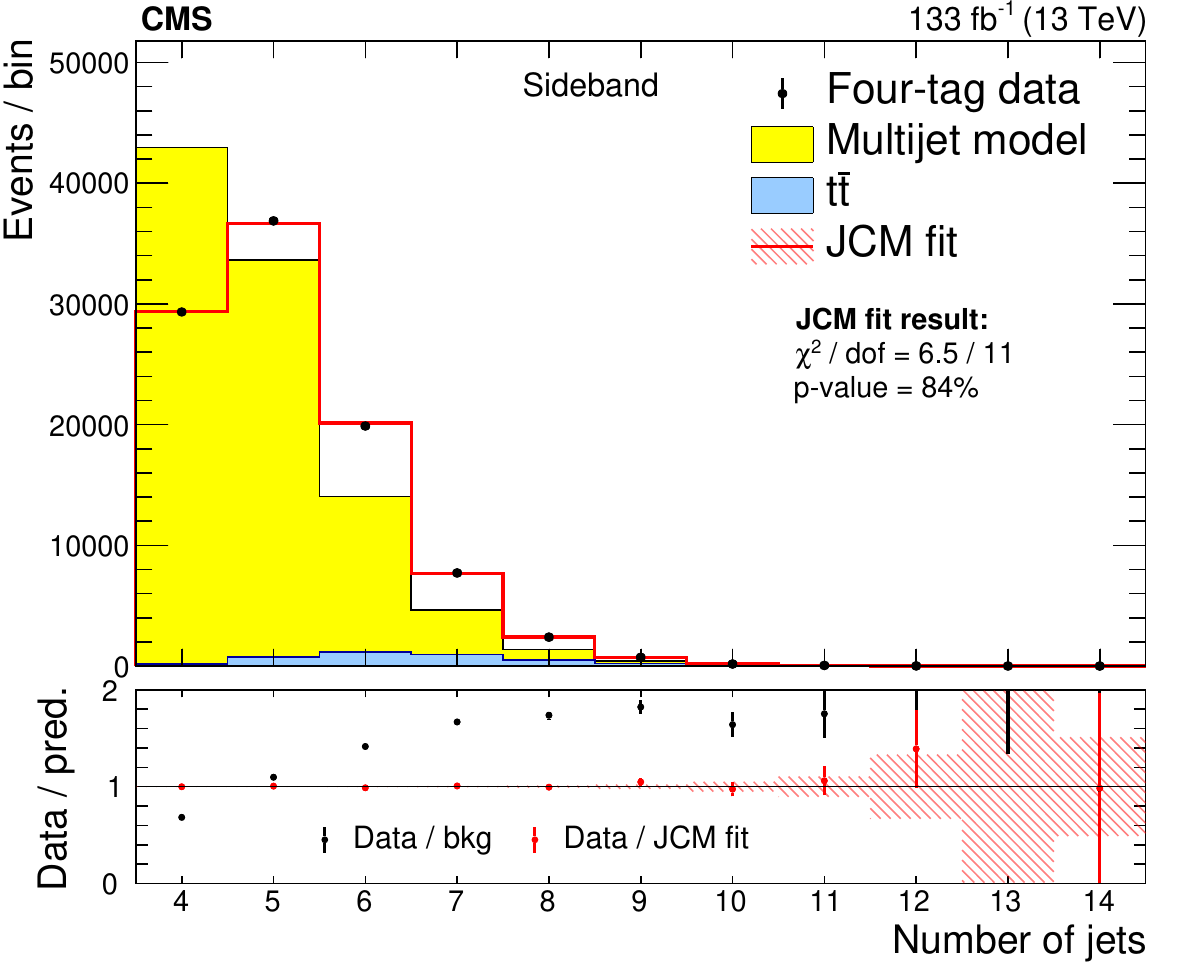}
      \includegraphics[width=0.49\textwidth]{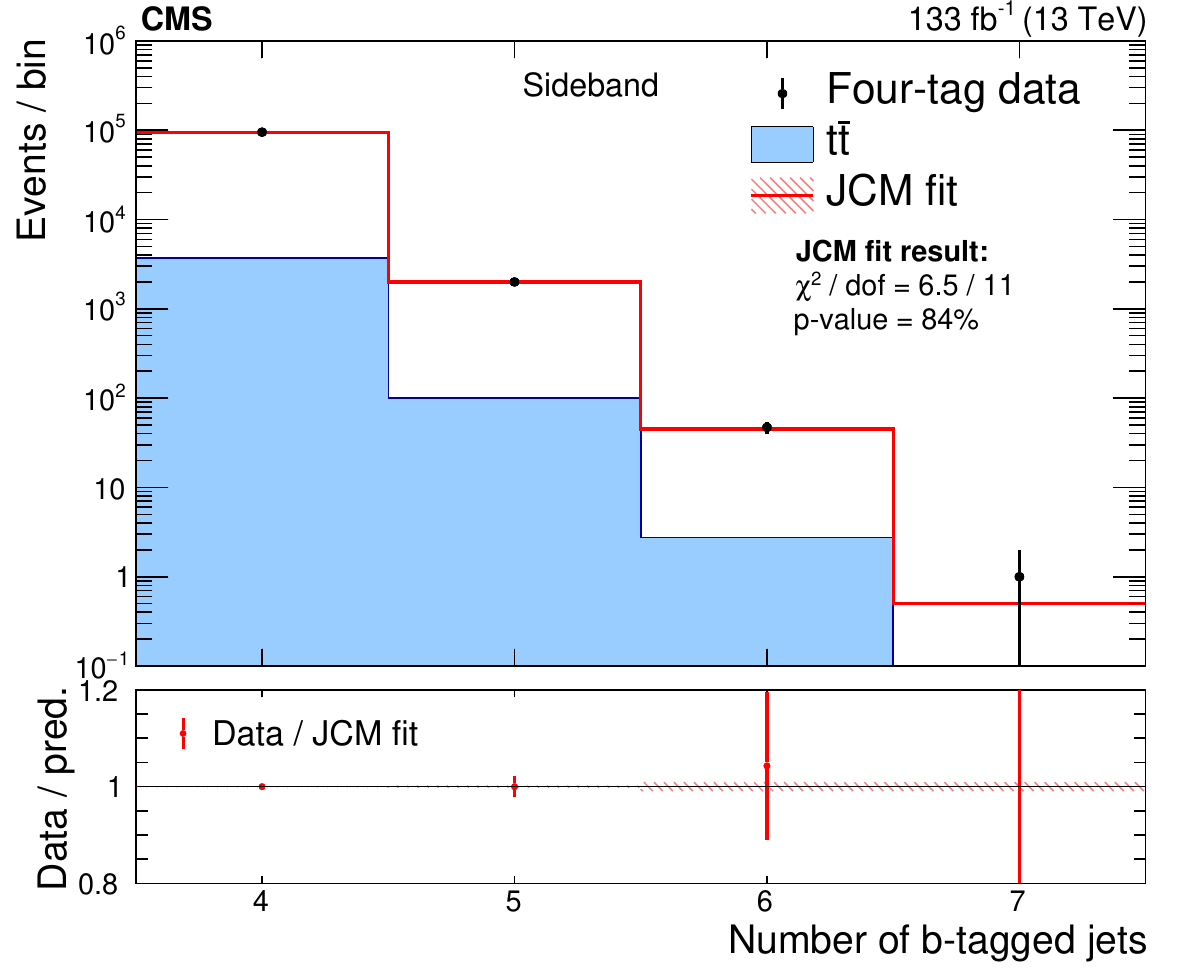}
  \caption{Jet (\cmsLeft) and \btagged jet (\cmsRight) multiplicity distributions in the SB region.
          The black data points show the observed four-tag data, the blue distribution the \ttbar simulation, and the yellow histogram the three-tag multijet prior to the JCM corrections.
          The red histogram shows the result of the fit to the JCM model. 
          The quality of the fit is given by the $\chi^2$ per degrees of freedom (dof) and corresponding $p$-value in the legend.
          The lower panels display the ratio of the data to the fit prediction.
          }
  \label{fig:JCM}
\end{figure}

The JCM parameters are determined by a combined fit to the jet and \btagged jet multiplicity distributions in the SB region.
Selected jets are required to satisfy the same kinematic requirements, $\pt > 40$\GeV and $\abs{\eta} < 2.4$, as the boson-candidate jets.
The jet and \btagged jet multiplicity distributions are shown in Fig.~\ref{fig:JCM} (\cmsLeft) and (\cmsRight), respectively, along with the results of the fit.
The black data points show the observed four-tag data; the yellow distribution displays the multijet background estimate prior to the JCM correction, and the blue distribution is from the \ttbar simulation.
The multijet background prior to the JCM correction is given by the \ttbar-subtracted three-tag data, normalized to the \ttbar-subtracted four-tag data.
This component is not shown on the \cmsRight panel as the three-tag data cannot be used to model the \btagged\ jet multiplicity.
The red histogram shows the result of the JCM fit, which provides a good description of both the jet and \btagged jet multiplicities.

\subsection{Kinematic reweighting}\label{sec:HCR_FvT}

\begin{figure}[hbt!]\centering
  \includegraphics[width=0.49\textwidth]{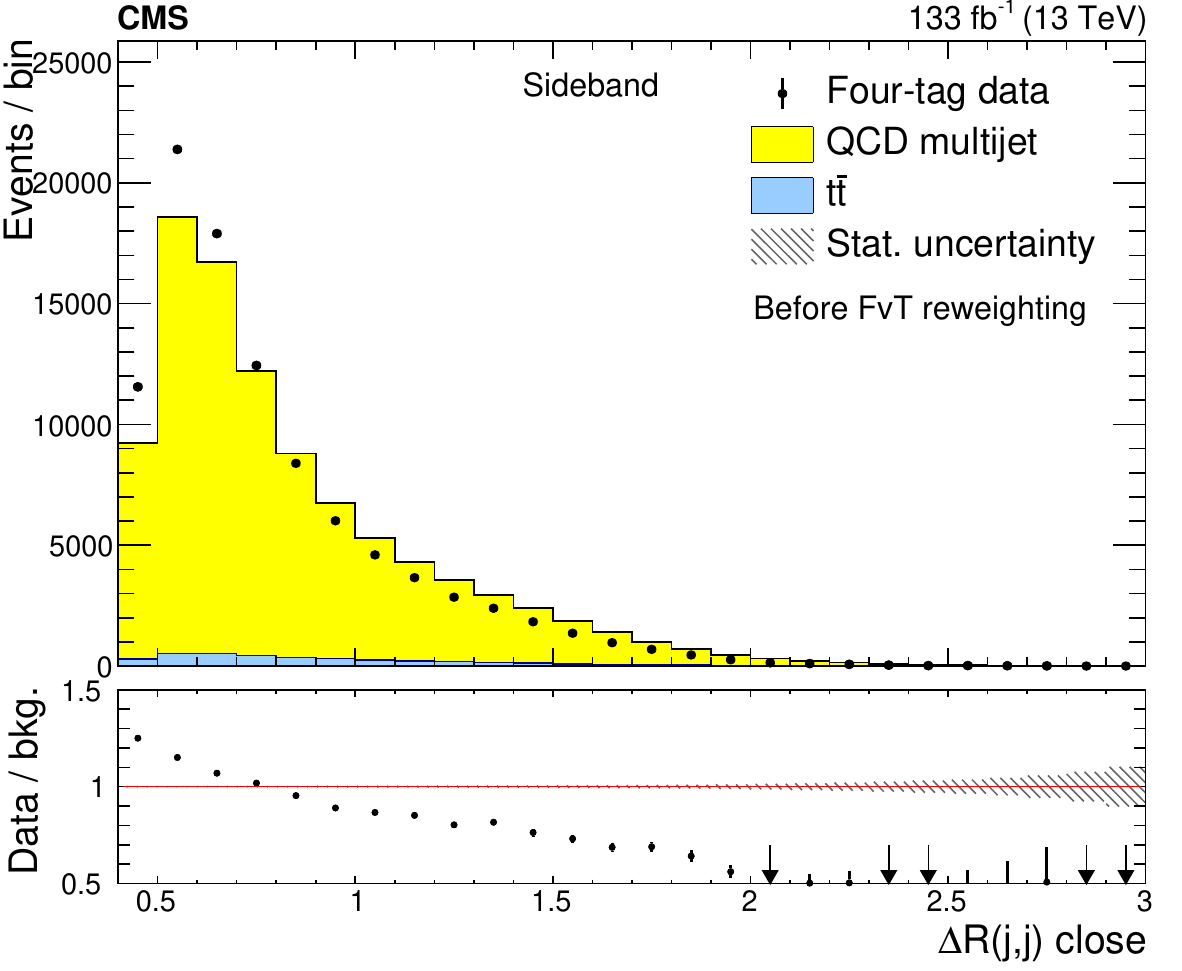}\hfill
  \includegraphics[width=0.49\textwidth]{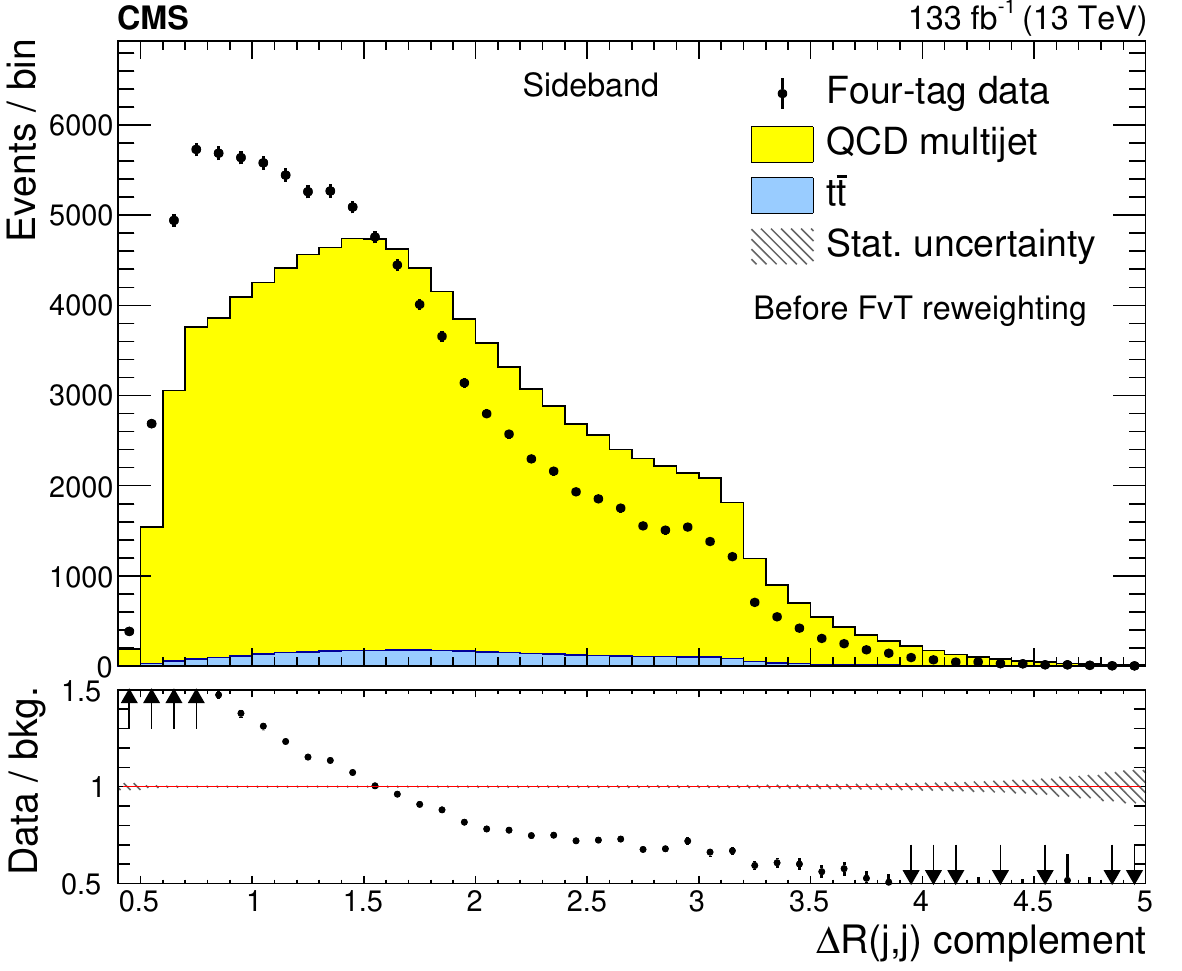}
  \caption{
  Distributions of $\Delta R(j, j)_{\text{close}}$ (\cmsLeft) and $\Delta R(j, j)_{\text{complement}}$ (\cmsRight).
  The four-tag SB events are shown by the points.
  The QCD multijet distribution  (yellow region) is from the three-tag SB sample after the JCM correction but before the FvT kinematic reweighting, and the \ttbar\ distribution (blue region) is from simulation.
  The lower panels display the ratio of the four-tag data to the total background, which is the sum of the QCD multijet and \ttbar\ distributions.
  The hatched area gives the statistical uncertainty in the background.
  }
  \label{fig:dR}
\end{figure}

\begin{figure}[hbt!]\centering
  \includegraphics[width=0.49\textwidth]{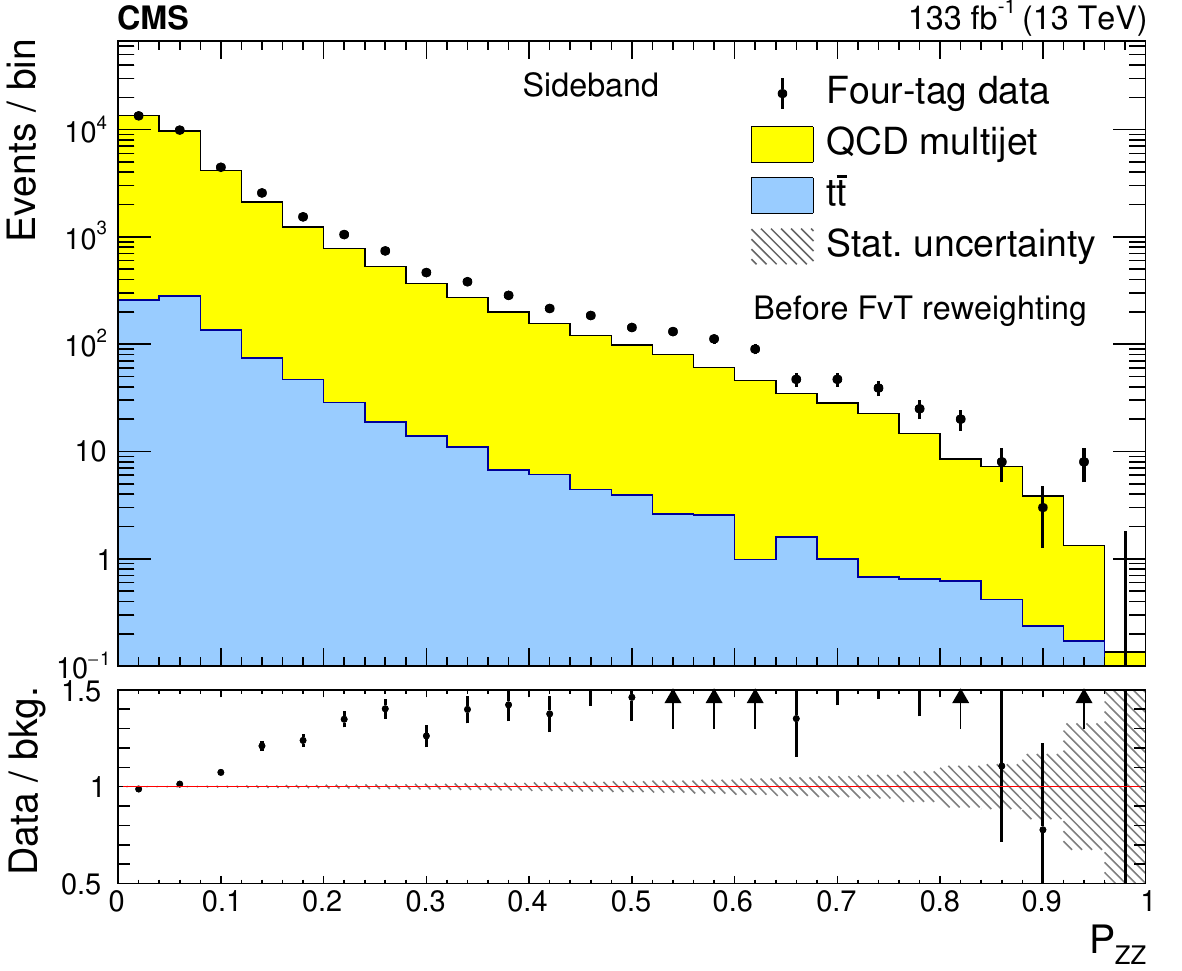}\hfill
  \includegraphics[width=0.49\textwidth]{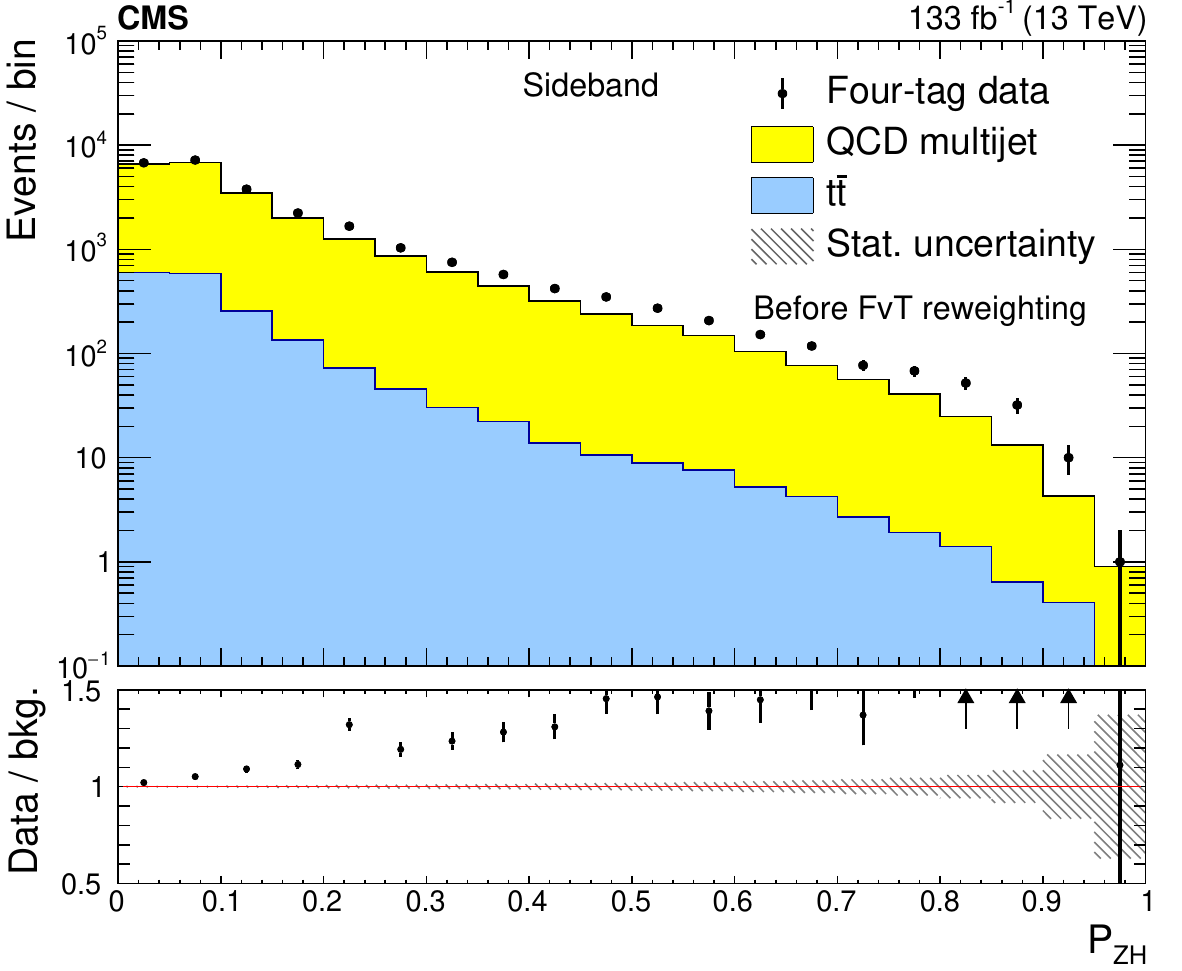}
  \caption{
  Distributions of the signal probabilities for \ZZ (\cmsLeft) and \ZH (\cmsRight) in the SB region, respectively.
  The four-tag SB events are shown by the points. 
  The QCD multijet distribution (yellow region) is from the three-tag SB sample after the JCM correction but before the FvT kinematic reweighting, and the \ttbar\ distribution (blue region) is from simulation.
  The lower panels display the ratio of the four-tag data to the total background, which is the sum of the QCD multijet and \ttbar\ distributions.
  The hatched area gives the statistical uncertainty in the background.
}
  \label{fig:SvB_ps}
\end{figure}

After correction with the JCM weights, differences remain between the three- and four-tag samples arising from the kinematic dependence of the \PQb tagging efficiency and from a different mixture of underlying scattering processes in the two selections. 
These differences are highlighted in Fig.~\ref{fig:dR}, which shows the opening angles $\Delta R(j, j)$ for dijet pairs, chosen such that the boson-candidate jets with the smallest opening angle form one pair (close dijet) and the remaining boson-candidate jets form the other (complement dijet).
The opening angles are shown for events in the SB region.
The multijet model, shown in yellow, includes the JCM weights.
The four-tag sample is dominated by gluon splitting to \bbbar\ from an underlying two-to-two scattering process.
This produces a topology with two \btagged dijets, each with a small \dR.
The three-tag sample contains this process in addition to a mixture of processes where the anti-b-tagged boson-candidate jet can be produced without gluon splitting, leading to a broader distribution of opening angles.
Significant differences between the three- and four-tag samples are also observed in other jet, dijet, and event-level distributions.
Figure~\ref{fig:SvB_ps} shows the modeling of the final discriminating variables --- the SvB signal probabilities for \ZZ ($\mathrm{P}_{\ZZ}$) and \ZH ($\mathrm{P}_{\ZH}$) --- in the SB region. 

The residual kinematic differences between the three- and four-tag samples are corrected using weights derived from a multivariate classifier with the HCR architecture.
This classifier, referred to as the FvT classifier, is trained with four classification targets: four-tag data, four-tag \ttbar\ simulation, three-tag data, and three-tag \ttbar\ simulation.
The FvT classifier is trained using data and the \ttbar simulation in the SB region.
The JCM weights are applied to the three-tag data and three-tag \ttbar simulation prior to training.
The output probabilities are used to reweight the three-tag data to the four-tag multijet background.
The FvT weights are given by
\begin{equation}
  \label{eqn:reweight}
  w_{\text{FvT}} = \frac{\pMfour}{\pDthree} \equiv \frac{\pDfour - \pTfour}{\pDthree}, 
\end{equation}
where \pDfour, \pTfour, and \pDthree\ are the class-assignment probabilities coming from the FvT classifier output,  and \ensuremath{\text{M}_{\text{4b}}} refers to the QCD multijet four-tag event. 
The final background model is given by the four-tag \ttbar simulation plus the three-tag data weighted by the product of the JCM and FvT weights.

\begin{figure}[hbtp]\centering
  \includegraphics[width=0.49\textwidth]{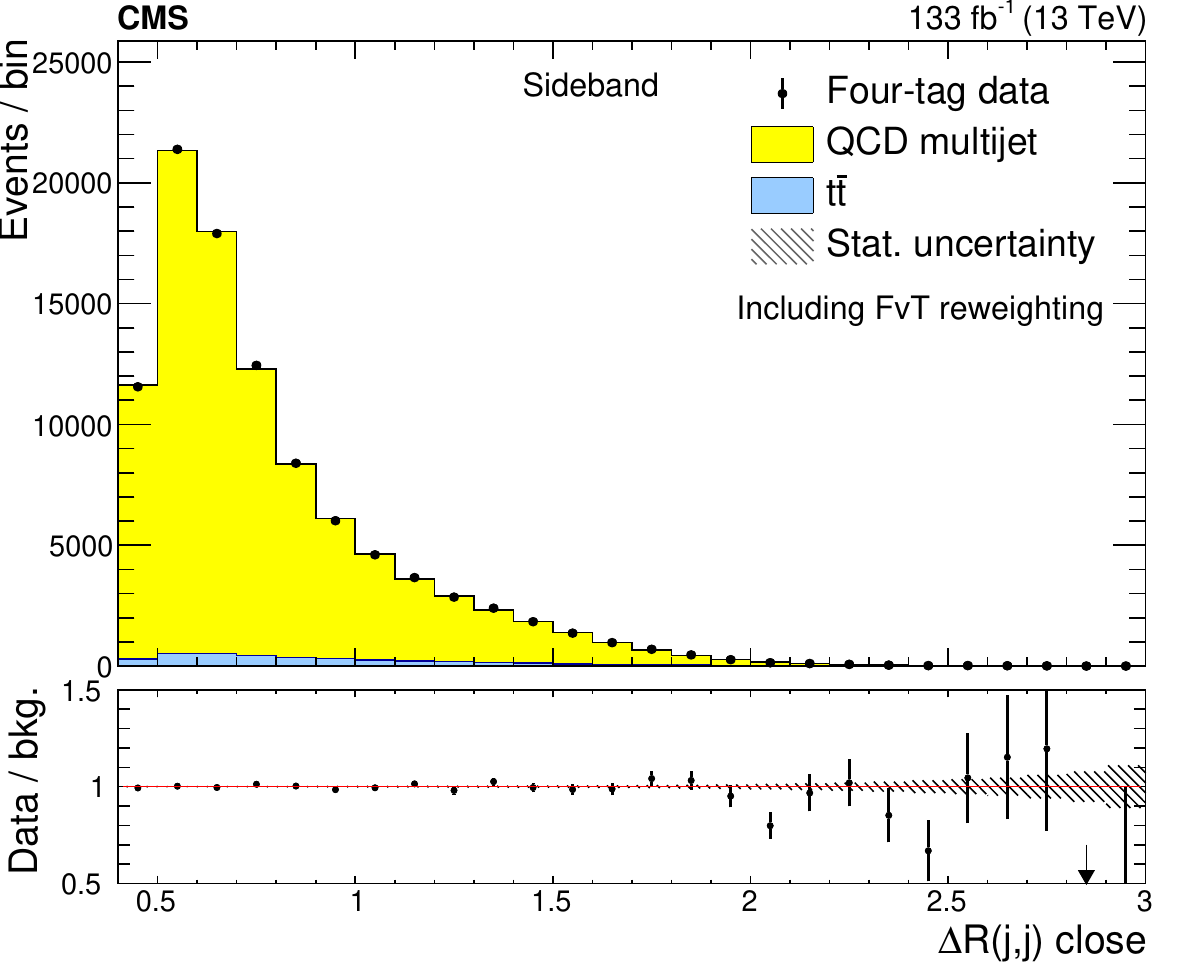}\hfill
  \includegraphics[width=0.49\textwidth]{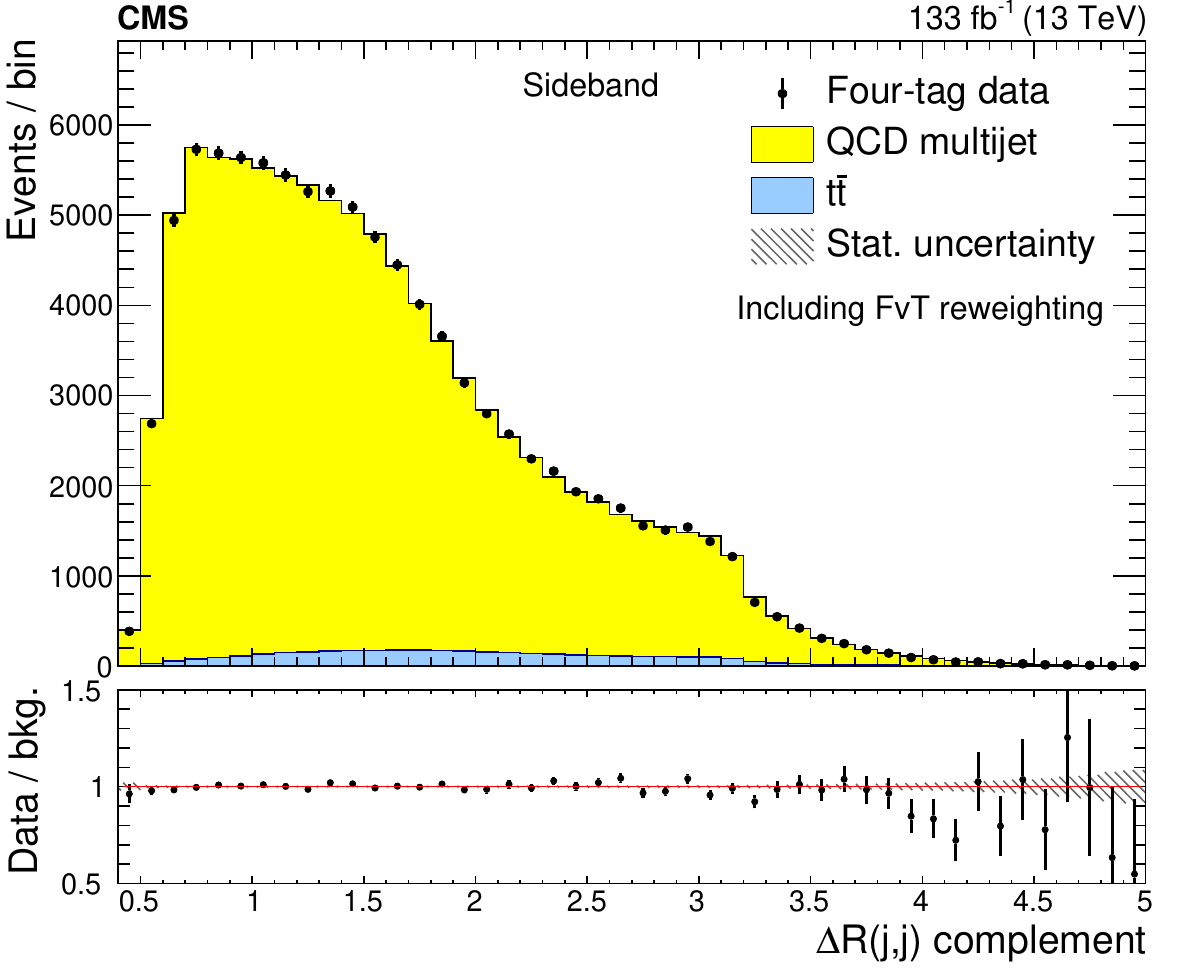}
  \caption{The \dR distributions shown in Figure~\ref{fig:dR} after including the FvT corrections to the QCD multijet prediction.
  }
  \label{fig:dR_reweight}
\end{figure}

\begin{figure}[hbtp]\centering
  \includegraphics[width=0.49\textwidth]{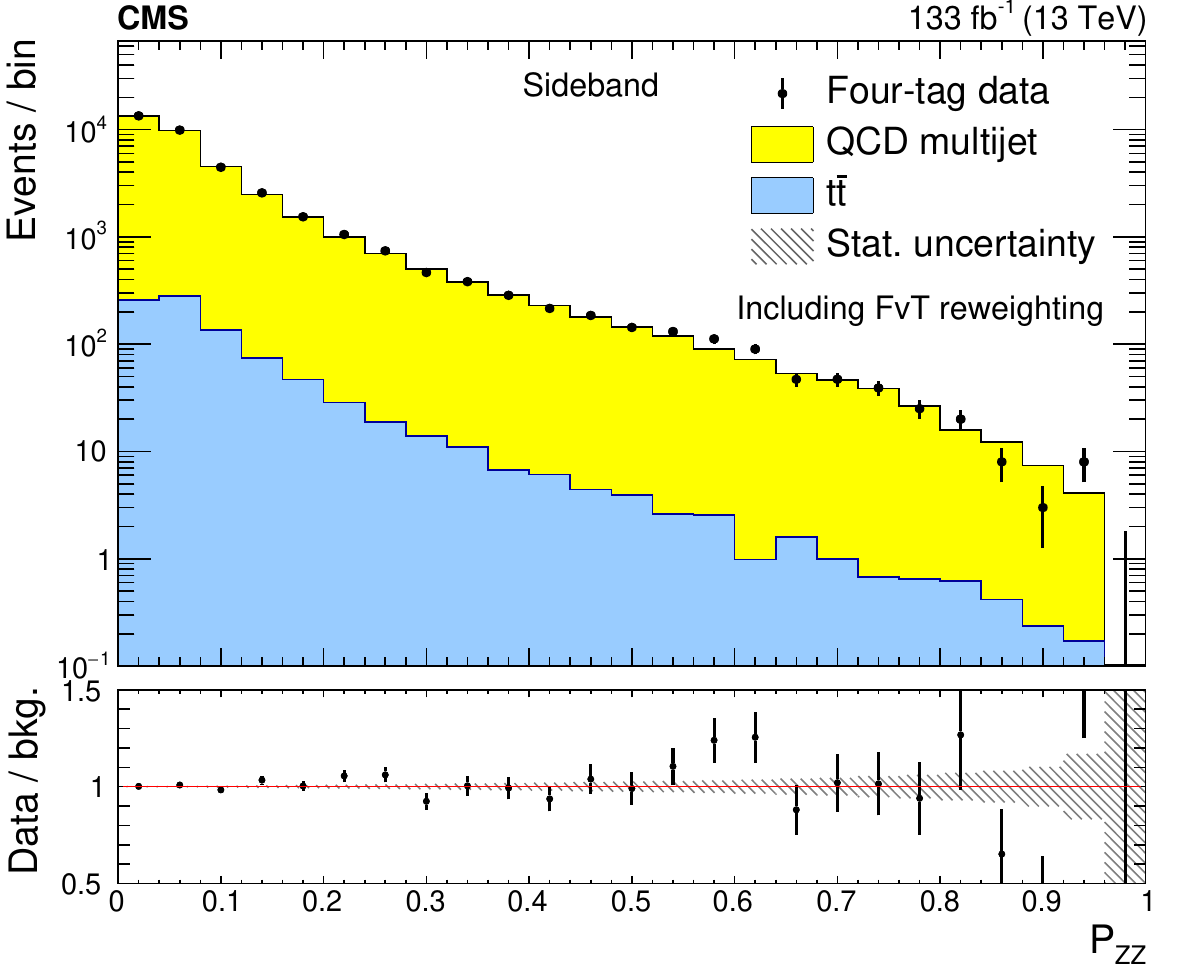}\hfill
  \includegraphics[width=0.49\textwidth]{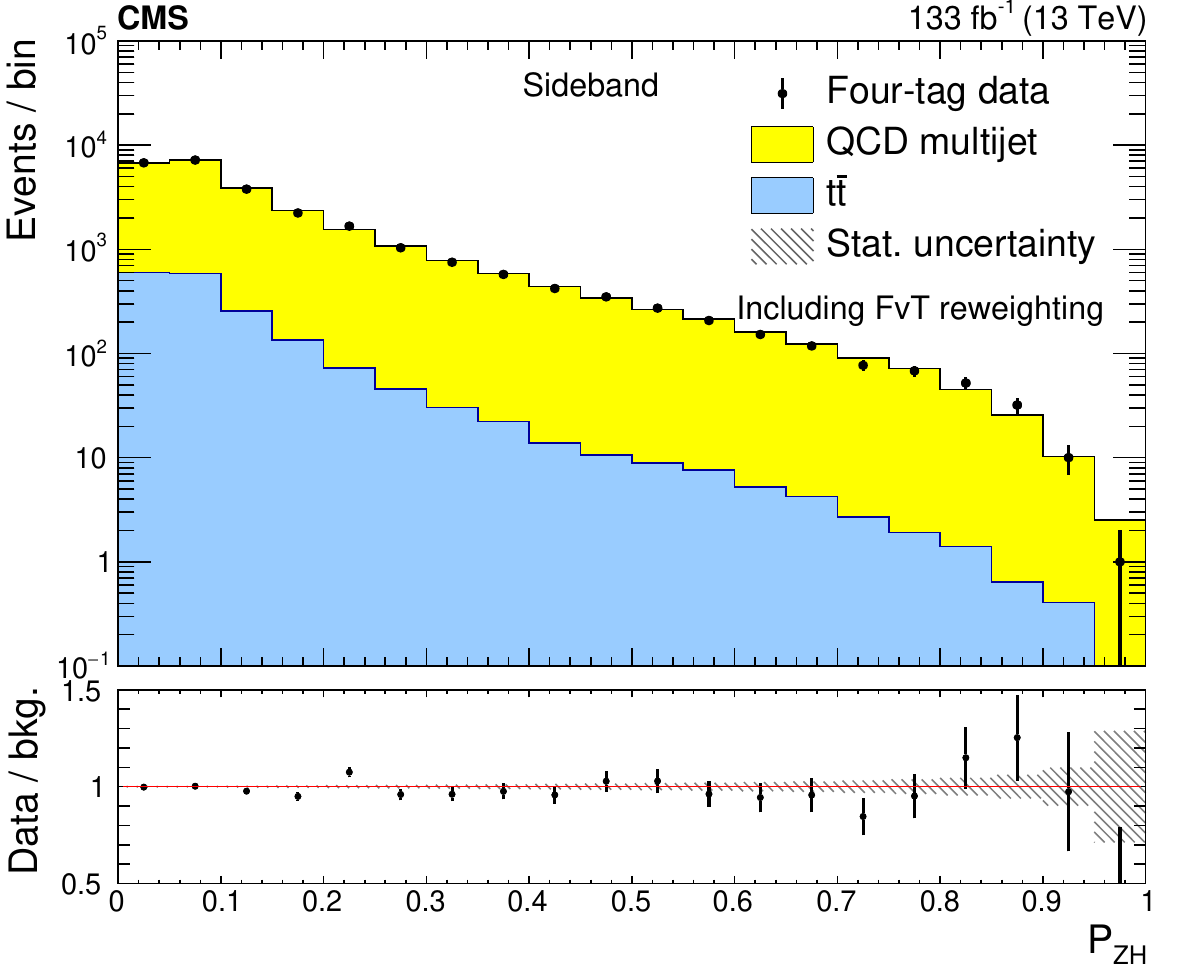}
  \caption{Distribution of signal probabilities for \ZZ (\cmsLeft) and \ZH (\cmsRight)  events in the SB region after including the FvT corrections to the QCD multijet prediction.}
  \label{fig:SvB_ps_reweight}
\end{figure}

Figure~\ref{fig:dR_reweight} shows the improvement in modeling of the close and complement candidate opening angles when the FvT weights are applied.
The FvT weights also correct the modeling of the other observed discrepancies in jet, dijet, and event-level distributions.
The impact of the FvT reweighting on the modeling of the SvB signal probabilities in the SB region is shown in Fig.~\ref{fig:SvB_ps_reweight}.

\section{Background validation}\label{sec:closureTest}

The accurate modeling of the SvB signal probabilities in Fig.~\ref{fig:SvB_ps_reweight} comes with two major caveats.
The first is that the FvT classifier is trained with the four-tag SB region data and thus has access to all of the relevant SvB information during training.
The final analysis in the SR requires an extrapolation of the background model to a different region of phase space.
This extrapolation is a significant source of systematic uncertainty that cannot be assessed in the region used to train the classifier.
The second major caveat is that the sensitivity in the SR is dominated by SvB signal probabilities above ${\approx}$ 0.9, for which there is little statistical power in the SB region.

These problems are addressed by validating the background model with synthetic data sets that allow the extrapolation of the background model to be tested precisely in regions of high signal probability.
Section~\ref{sec:mixing} describes the construction of the synthetic data sets, and Section~\ref{sec:bkgSys} describes how it is used to assess the systematic uncertainties in the background model.

\subsection{Synthetic data sets from hemisphere mixing}\label{sec:mixing}

A synthetic data set is generated by splitting individual events into hemispheres and then combining similar hemispheres from different events.
This mixing procedure suppresses correlations among the jet four-vectors due to the presence of a signal, while preserving the kinematic distributions of the 4\PQb background.
The relevant event-level correlations in the background --- which primarily arise from gluon splitting in an underlying two-to-two scattering process --- are captured in the correlation between hemispheres and are independent of the dijet substructure within the hemispheres.

The mixing algorithm is based on the technique developed in a previous CMS $\HH\to4\PQb$ analysis~\cite{CMS:2018sxu}.
The first step involves creating a collection of hemispheres (hemisphere library) from events in the four-tag data set.
Each event is split into two hemispheres using the plane orthogonal to the transverse thrust axis~\cite{CMS:2018sxu}, which is chosen based on the assumption that it is a good proxy for the initial gluon directions in the underlying scattering process.
Jets on one side of the plane are assigned to one hemisphere, those on the other side are assigned to the other hemisphere.
Four variables are computed using the sum of the four-vectors of all the jets in that hemisphere: the invariant mass, the longitudinal momentum, and the transverse momentum perpendicular and parallel to the transverse thrust axis.
The jet and \bjet\ multiplicities are also computed for each hemisphere.
The library is created with events that pass the jet kinematic requirements but before the dijet invariant mass requirement given in Eq. (\ref{eqn:SB}) is applied.

The mixing is performed in a second pass over the data.
Events are split into hemispheres as before, and the corresponding hemisphere summary variables are calculated.
Each hemisphere in the event is replaced with its nearest neighbor in the hemisphere library.
Nearest neighbors are defined as hemispheres with the same jet and \bjet\ multiplicities that minimize the distance between hemispheres, defined as
\begin{equation}\label{eq:metric}
 d(h_i,h_j) = \sqrt{\sum_{k} \left(\frac{v_{k}(h_i) - v_{k}(h_j)}{\sigma_{v_{k}}}\right)^2},
\end{equation}
where $h_i$ and $h_j$ are two hemispheres and the sum runs over the four hemisphere summary variables $v_k$ and  $\sigma_{v_{k}}$ is the root mean square of the corresponding variable calculated from all hemispheres in the library.
During the nearest neighbor replacement a check is made that the matching hemispheres do not come from the same event.
Finally, the nearest neighbor hemispheres are rotated in the azimuthal angle $\phi$ to match the direction of the transverse thrust axis of the input event.

\begin{figure*}\centering
\includegraphics[width=0.98\textwidth]{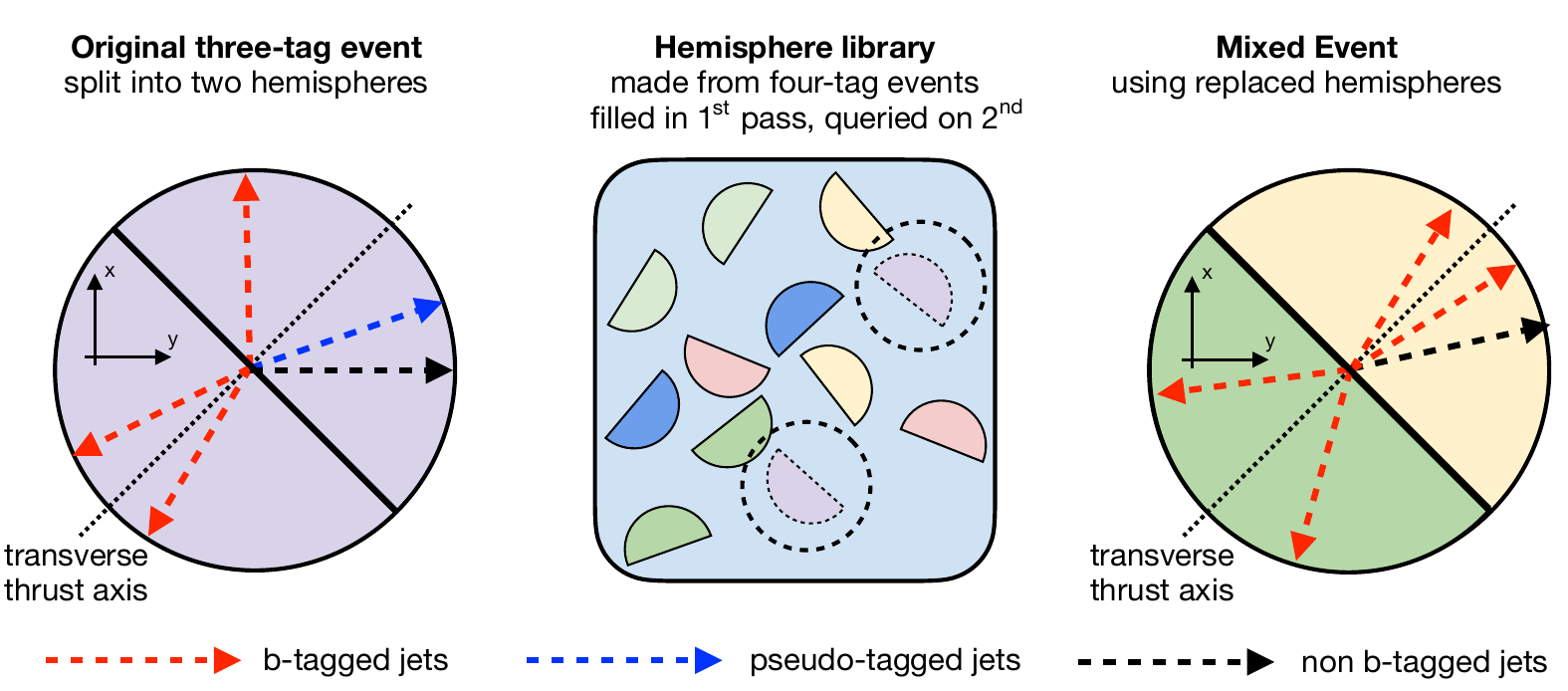}\hfill
\caption{
An illustration of the hemisphere mixing procedure, adapted from Ref.~\cite{CMS:2018sxu}.
Three-tag events are divided into two halves by cutting along the axis perpendicular to the transverse thrust axis.
In a preliminary step, each event in the four-tag data set is split into two hemispheres that are collected in a library of hemispheres.
Once the library is created, each three-tag event is used as a basis for creating a synthetic event.
These are constructed by picking the two hemispheres from the library that are most similar to the hemispheres making up the original event.
}
\label{fig:hemiCartoon}
\end{figure*}

This analysis introduces two important improvements to the  mixing strategy.
The first is the use of the three-tag data set in mixing. 
Four-tag events are used to create the hemisphere library, however, the three-tag data set is used in the second pass when creating the mixed data sample. 
In the three-tag sample, the pseudo-tagged jets are treated as \PQb tagged when matching hemispheres; 
this ensures that events in the resulting mixed data sample all have four \btagged\ jets.
An illustration of the mixing procedure is given in Fig.~\ref{fig:hemiCartoon}.
Mixing the three-tag data eliminates a potential bias from signal contamination and allows for the construction of a synthetic data sample that is fifteen times larger than the four-tag data sample. 

The other important improvement to the mixing algorithm is the introduction of corrections accounting for the \ttbar background contamination.
The presence of the \ttbar events complicates the mixing strategy, since these events have significant hemisphere-level correlations coming from the decays of the top quarks.
Mixing a \ttbar\ hemisphere with a QCD multijet hemisphere would significantly distort the \ttbar\ background.
The mixed \ttbar\ events are thus not expected to provide a good model of the unmixed \ttbar\ background.

The \ttbar\ background needs to be subtracted from the hemisphere library and the three-tag data set being mixed --- not from projections into histograms, as is more typical in high energy physics.
These \ttbar\ contributions are subtracted statistically using event weights derived from a multivariate classifier with the HCR architecture.
This classifier is trained with two classes, data and \ttbar\ simulation, separately on events in the three- and four-tag data set. 
The outputs from the classifier are used to determine the probability \pM\ that an event in the three- or four-tag data sample is a multijet event.
When constructing the mixed data sample, an input three-tag event is accepted for mixing with probability \pM.
Similarly, nearest-neighbor hemispheres are accepted as replacements with probability \pM, as calculated in their corresponding four-tag event.
When a hemisphere is rejected, the next-nearest neighbor is considered in turn.

The size of the four-tag data sample in the SB region, used to train the FvT classifier, is a fundamental limitation of the background determination procedure.
When validating the background, it is critical that the synthetic data have the same statistical power as the nominal four-tag sample. 
To ensure this, the mixed data are subsampled to match the size of the QCD multijet background in the four-tag sample.
The relative size of the mixed data sample allows fifteen independent subsamples to be created.
The number of subsamples is somewhat smaller than the relative size of the three-tag dataset to avoid using duplicate events with large jet multiplicity.
Events from the four-tag \ttbar\ simulation are added to each subsample according to the expected \ttbar\ yield.
These fifteen synthetic data sets are referred to as mixed models.

\begin{figure*}\centering
\includegraphics[width=0.49\textwidth]{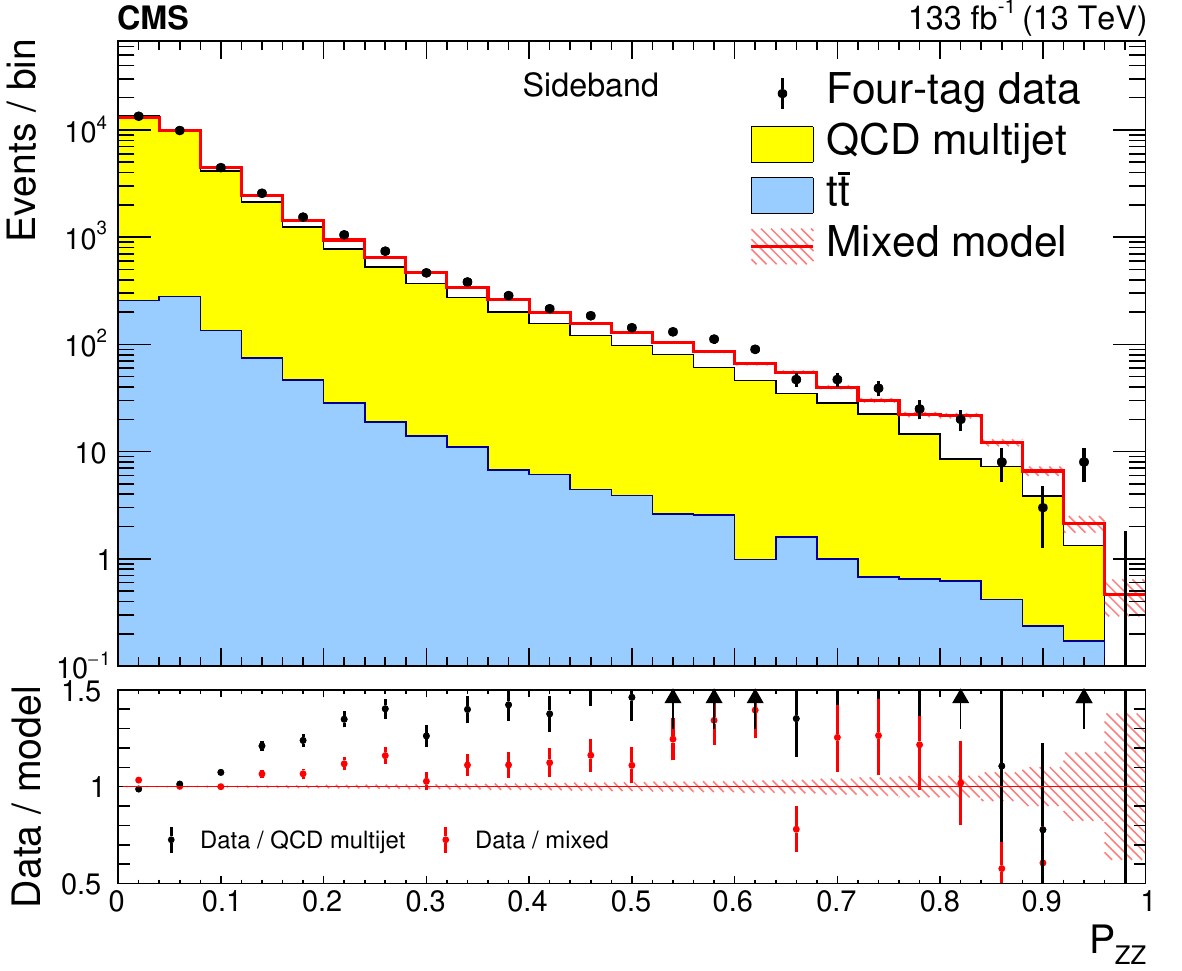}\hfill
\includegraphics[width=0.49\textwidth]{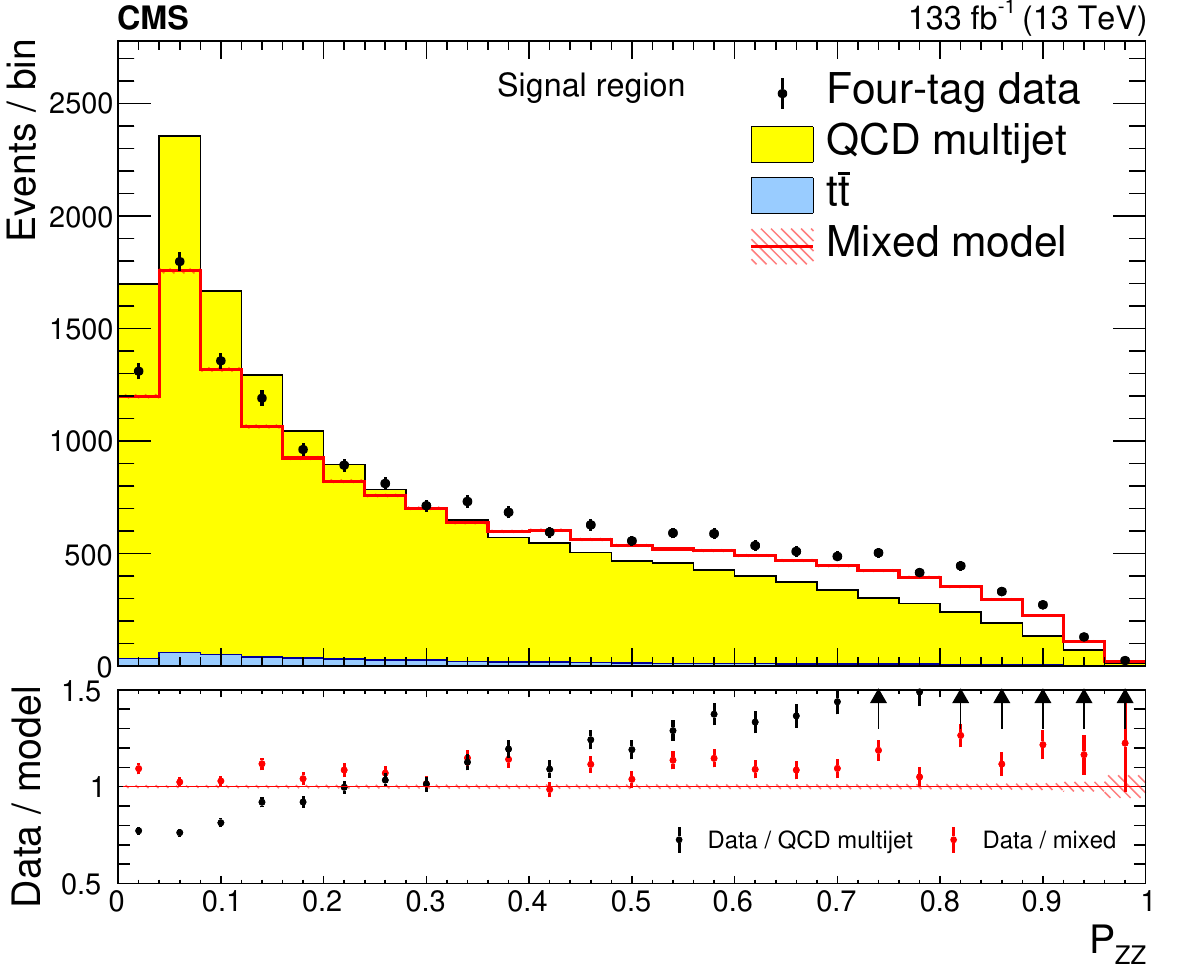}\vspace{0.1in}
\includegraphics[width=0.49\textwidth]{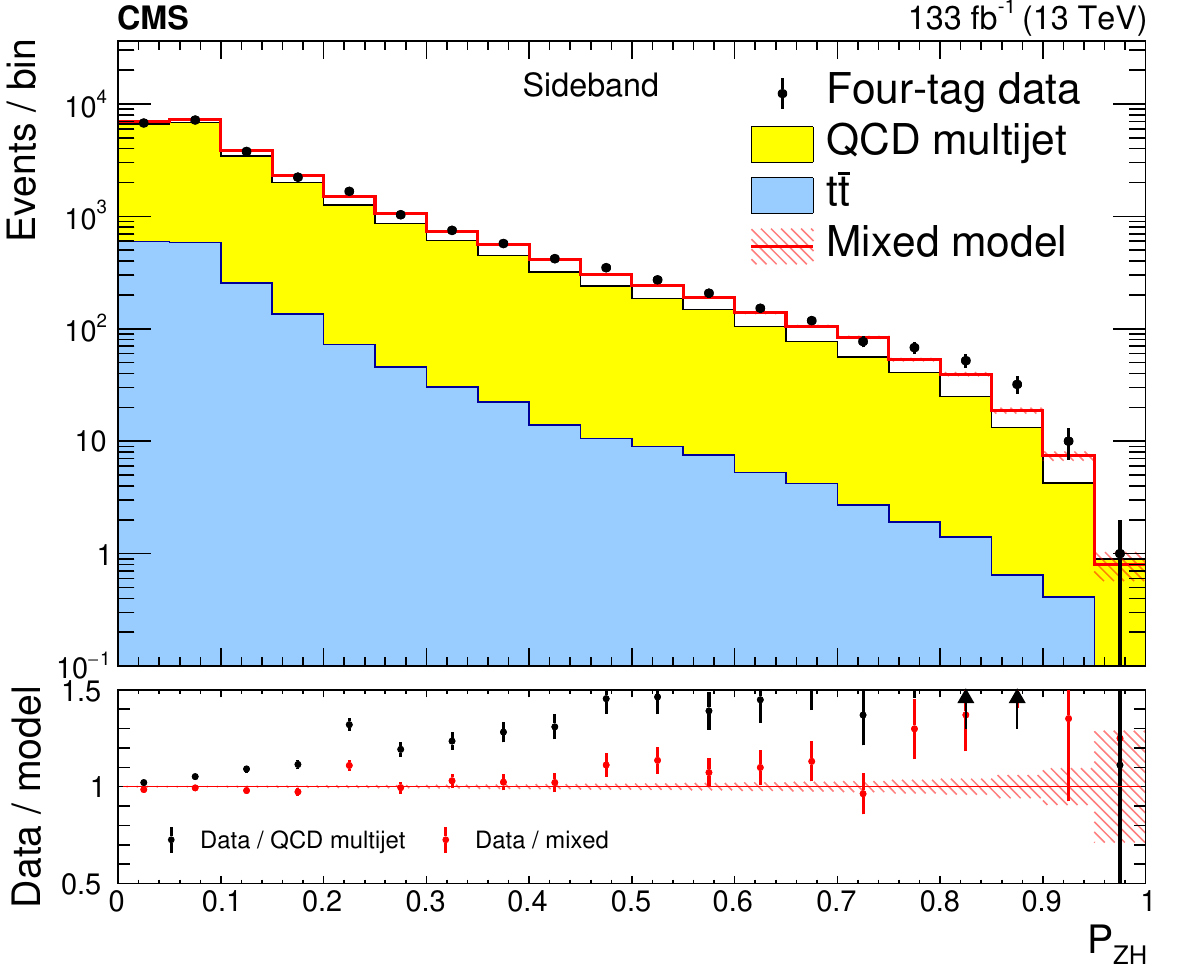} 
\includegraphics[width=0.49\textwidth]{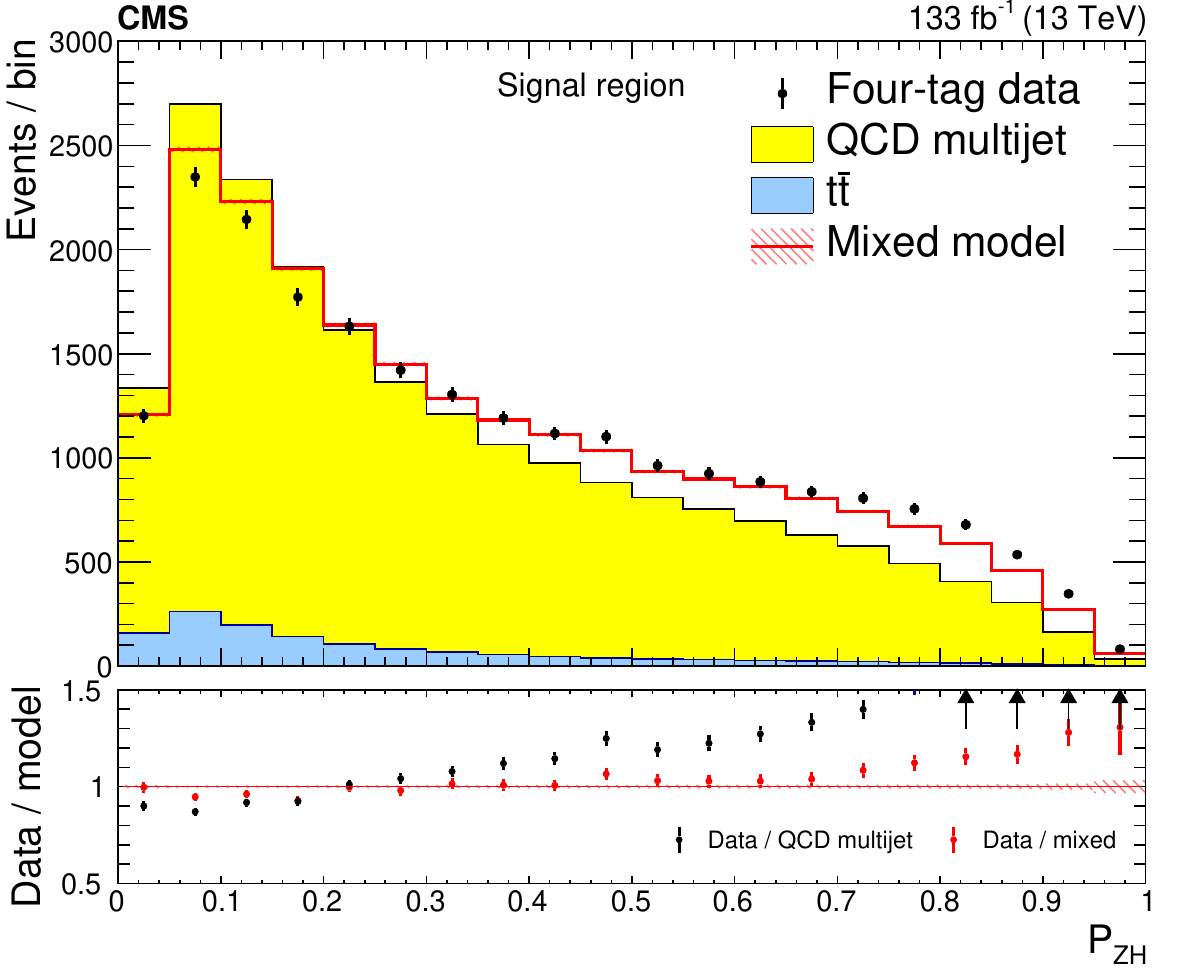}
\caption{
        Distribution of signal probabilities for \ZZ (upper row) and \ZH (lower row) events in the sideband (left) and signal regions (right).
        The four-tag events are shown by the points.
        The QCD multijet distribution before the FvT corrections is given by the yellow region, and the simulated \ttbar\ distribution by the blue area.
        The average of the mixed models (red) provides a high-event-count proxy of the 4\PQb background (black) that allows the extrapolation of the background model to be tested precisely.
        The lower panels display the ratio of the four-tag data to the average of the mixed models (red) and to the QCD multijet distribution (black).
        }
\label{fig:summaryOfClosure}
\end{figure*}

The advantages of using the mixed models to validate the background can be seen in Fig.~\ref{fig:summaryOfClosure}.
The \ZZ (\ZH) SvB signal probabilities are shown in the upper (lower) figures in the SB region, left, and in the SR, right.
The four-tag data are shown by the black points.
The three-tag data distribution before applying the kinematic FvT corrections, from which the background prediction is extrapolated, is shown in yellow.
The average of the mixed models, shown in red, provides a high-event-count proxy for the four-tag background that can be used to validate the background prediction in regions of high signal probability.

\subsection{Background model uncertainties}\label{sec:bkgSys}

The background procedure, as described in Section~\ref{sec:background}, is repeated by treating the mixed models as four-tag data.
The procedure is performed separately using each of the fifteen mixed models.
Differences among the fifteen QCD multijet predictions and the comparison of the predicted background to the observed SR yield are used to assign systematic uncertainties in the nominal background model.
These systematic uncertainties are assessed in three steps.
First, the differences among the fifteen QCD multijet predictions are quantified.
These differences arise from the finite size of the data sample in the SB region used to train the FvT classifier. 
The second step measures the systematic uncertainty from extrapolating to the SR, by comparing the background predictions to the observed yields in the mixed models.
The final step checks if biases in the background model can mimic a signal.
This process is carried out independently for the final \ZZ and \ZH selections.

The differences in the background models are quantified by comparing the QCD multijet predictions of each mixed model in the SR to their average.
These differences are parameterized by a set of orthogonal Fourier basis functions added to the background predictions.
Each of the multijet predictions, with unconstrained coefficients for the basis function corrections, are fit separately to the average.
Basis functions with increasing frequency components are added until the pulls of adjacent SvB signal probability bins from all the mixed-model predictions are consistent with being uncorrelated.
A pull is defined as the difference between the observed and expected values, divided by the uncertainty in the difference.
A systematic uncertainty in each coefficient is assigned based on the root mean square of the fitted values.
This uncertainty accounts for the expected variance of a single background prediction due to the finite size of the data sample in the SB region.

This procedure is carried out separately for the \ZZ and \ZH signal probabilities.
Five (four) basis functions, with uncertainties in the basis-function coefficients of up to 3\%, are needed to characterize the differences among the \ZZ (\ZH) background predictions.

The systematic uncertainty from extrapolating the background prediction is evaluated by comparing the SR predictions to the observed yields in the mixed models.
A combined background model is fit to the average of the observed SR yields.
Averaging the fifteen mixed models improves the precision with which the extrapolation uncertainty can be determined.
The combined background model consists of the estimated \ttbar, the average of the QCD multijet predictions, and the basis-function corrections determined above.
The coefficients of the basis functions are treated as nuisance parameters constrained using the systematic uncertainties assigned in the previous step.

The extrapolation uncertainty is quantified using the basis-function coefficients determined from fitting the mixed models.
The fit is repeated, sequentially removing constraints on the nuisance parameters, until the fit has a $p$-value greater than 5\% and an F-test~\cite{FTest} does not prefer more unconstrained basis function coefficients.
Nonzero fitted coefficients represent a systematic difference between the predicted and observed background.
Systematic uncertainties in the background extrapolation are assigned by adding the magnitude of the fitted coefficients in quadrature with their uncertainty.
These extrapolation uncertainties are treated as uncorrelated from the variance uncertainties assigned in the previous step.

\begin{figure}[hbt!]
  \includegraphics[width=0.49\textwidth]{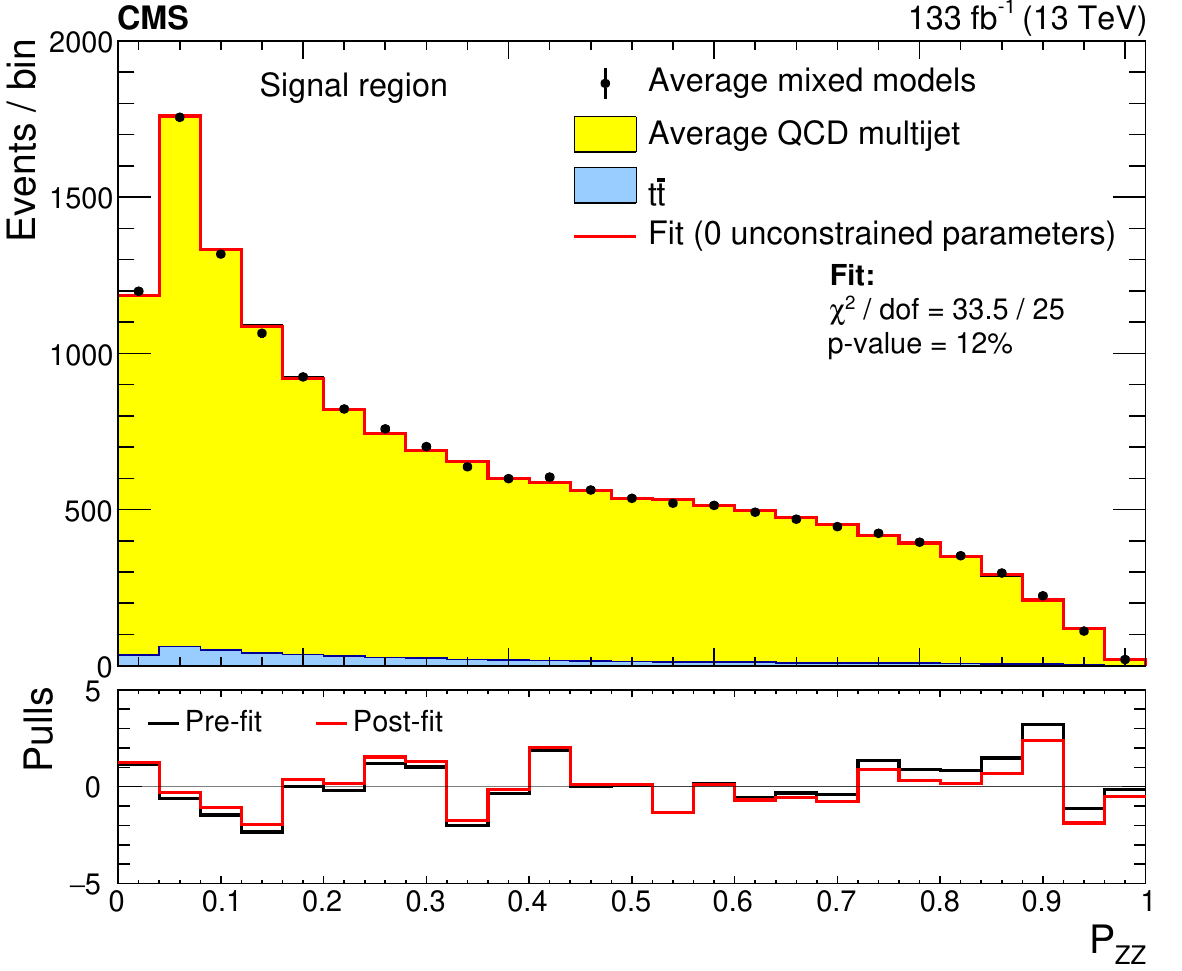}\hfill
  \includegraphics[width=0.49\textwidth]{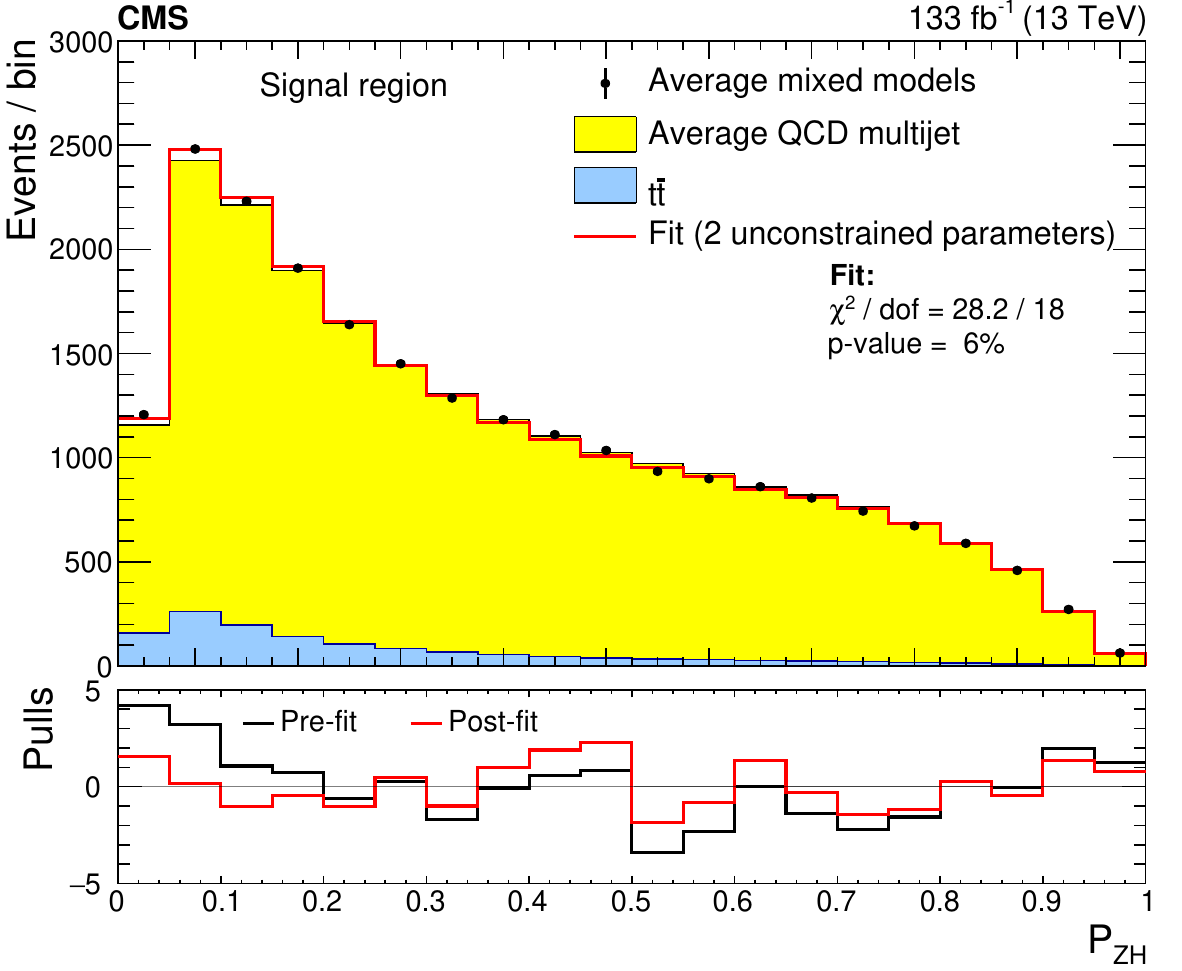}\hfill
  \caption{
     The distributions of the \ZZ (\cmsLeft) and \ZH (\cmsRight) signal probabilities.
       The black data points show the average of the mixed models.
       The yellow and blue distributions show the average of the QCD multijet models and the \ttbar\ simulation, respectively.
     The red histogram displays the post-fit results of the data fit to the background model.
     The \ZZ channel data distribution is fit with all five basic coefficients constrained, while the \ZH channel distribution has two of the four coefficients unconstrained.
     The lower panels give the pre- (blue) and post-fit (red) pulls.
  }
  \label{fig:bias_zz_zh}
\end{figure}

Figure~\ref{fig:bias_zz_zh} illustrates the process in determining the extrapolation uncertainty in the \ZZ (left) and \ZH (right) SRs.
The upper panels compare the mixed model SvB signal probability in the SR, given by the black points, to the pre-fit background prediction, shown as stacked yellow and blue histograms, and the post-fit background prediction, shown in red.
The lower panels show the pulls.
For the \ZZ SR, none of the fit parameters need to be unconstrained to satisfy the goodness-of-fit or the F-test criteria.
For the \ZH SR, the goodness-of-fit criteria is not satisfied until two parameters are unconstrained, at which point the F-test criterion is also satisfied.
The extrapolation uncertainty is ${\lesssim}\ 1\%$ for most parameters and at most ${\approx}\ 3\% ({\approx}\ 5\%)$ in the \ZZ (\ZH) region.

The mixed models can also be used to test if biases in the background model can mimic the signature of a signal, a possibility to which many analyses with backgrounds estimated from control samples in data are blind.
To assess the risk of fitting a spurious signal, the fit to the averaged mixed models in the second step is repeated with and without an unconstrained signal template.
For this test, the coefficients of the basis functions in the background model are constrained with the systematic uncertainties assigned in the previous two steps.
An F-test is performed that compares the background-only model to the model including an unconstrained signal template.
In both SRs, it is found that allowing for a spurious signal does not lead to a significant improvement in the model fit, therefore no additional systematic uncertainty is assigned.

\section{Systematic uncertainties}\label{sec:systematics}

A maximum likelihood fit to the four-tag data is performed on the distribution  of the SvB signal probability simultaneously in the \ZZ and \ZH SRs.
Systematic uncertainties are treated as nuisance parameters with either Gaussian (shape uncertainties) or log-normal (normalization uncertainties) function priors included in the likelihood function.
All systematic uncertainties are considered as shape uncertainties with the exception of the luminosity, predicted signal cross section, and branching fraction uncertainties, which are treated as normalization uncertainties.

Table~\ref{tab:systematics} summarizes the impact of different sources of uncertainty in the \ZZ and \ZH signal sensitivity.
The table shows the relative contributions of the various sources of uncertainty in the measured signal strength, quoted in terms of a percentage of the total uncertainty.
The contributions from the leading sources of uncertainty -- background modeling, \btagging, and jet energy scale and resolution -- are listed separately.
The remaining uncertainties, described below, are included in the row labeled ``Others''.
The statistical uncertainty accounts for over half of the total uncertainty, while the remaining uncertainty is primarily due to experimental uncertainties in the background model and the \btagging\ efficiency.

\begin{table}
\topcaption{Summary of the relative uncertainties form the various sources in the measured signal strength, expressed as a percentage of the total uncertainty for the \ZZ and \ZH channels.
            The two uncertainties coming from the background modeling are given separately in parentheses, as well as their sum.
            The total systematic uncertainties shown include the effects of correlations.
            }\label{tab:systematics}
\renewcommand{\arraystretch}{1.2}
\centering
\begin{tabular}{lcc}
 Source                     & \ZZ  & \ZH   \\
\hline
 Statistical uncertainty          &  75        &  77 \\
Total systematic uncertainty   &  67        &  64 \\
Background model         &   61  &  56  \\
\hspace*{1.5cm}    (Variance)                   &   (46)    &  (46)  \\
\hspace*{1.5cm}    (Extrapolation)              &   (40)    &  (33)  \\
\hspace*{0.5cm}  b tagging                  &   9   &  17  \\
\hspace*{0.5cm}  Jet energy scale and resolution   &   9   &  5  \\
\hspace*{0.5cm}     Others                  &   24  & 24  \\
\end{tabular}
\end{table}

The uncertainties in the background model are described in Section~\ref{sec:closureTest}.
Similar results are obtained when characterizing the shape differences using a Fourier or a shifted Legendre polynomial basis.
Despite being determined with a relative error of a few percent, the uncertainty in the background prediction accounts for ${\approx}60\%$ of the total uncertainty in the measured signal strengths.

The efficiency of the \btagging\ requirement in the simulated samples is corrected to match the efficiency measured in data~\cite{CMS:2017wtu}.
These corrections, along with their corresponding uncertainties, are determined in bins of jet \pt, $\eta$, and the \textsc{DeepJet} \btagging\ score.
The largest uncertainties in the \btagging\ efficiency arise from contamination of light-flavor jets in heavy-flavor control regions.
To evaluate the impact of the \btagging\ uncertainties, the per-jet uncertainty in the \btagging\ corrections is propagated to the final SvB distribution.
The \btagging\ uncertainties result in a roughly flat ${\pm}20\%$ variation in signal yield, with no significant variations in the SvB shape, and contribute 10--20\% of the total uncertainty.

Uncertainties in the modeling of the jet energy scale and resolution, and the \bjet\ energy scale correction in the simulation are estimated by propagating variations of the calibrations~\cite{Khachatryan:2016kdb} to the final SvB discriminant distributions.
These variations change the reconstructed energy and direction of simulated jets and can thus result in event migration across regions and signal probability bins.
The combined jet energy and resolution uncertainties contribute 5--10\% of the total uncertainty.

The efficiency of the trigger requirement in the simulated samples is adjusted to match the efficiency measured in data.
The efficiencies for \bjets to pass the various trigger thresholds, based on their \pt and \btagging\ scores, are measured in a \ttbar sample where both top quarks decay leptonically.
Systematic uncertainties on the measured trigger efficiency are evaluated and applied to the expected signal yield.
The largest trigger uncertainty comes from the calculation of per-event trigger efficiencies using the measured per-jet efficiencies.
The total uncertainty in the trigger efficiency is estimated to be ${\approx}\ 5\%$ for both the \ZZ and \ZH signals.

The total uncertainty in the \ZZ (\ZH) cross section prediction is 6.6\% (4.1\%).
These uncertainties include the effects of varying the renormalization and factorization scales and the parton distribution function (PDF) of the proton.
The uncertainty from the choice of the factorization and renormalization scales in the calculation of the matrix element for the hard-scattering process is estimated by varying each scale by factors of 0.5 and 2, excluding anticorrelated combinations, to calculate the envelope around the central value.
In order to estimate the impact on the results due to the uncertainty on the proton PDF, event weights corresponding to the different set of NNPDF~\cite{NNPDF:2014otw} replicas are applied to the simulation.
The uncertainty in the \PH $\to$ \bb branching fraction is  $\pm$1.3\%~\cite{deFlorian:2016spz}.

The uncertainty in the total integrated luminosity for each data set has been measured in Refs.~\cite{CMS-LUM-17-003,CMS-PAS-LUM-17-004,CMS-PAS-LUM-18-002}.
A correlation scheme is used for the three sets of uncertainties based on correlated features in calibration methods, measurements, and data sets, resulting in an uncertainty of 1.6\% for the full data set.

The systematic uncertainty in the signal yields and distributions due to pileup is found to be negligibly small.

\section{Results}\label{sec:results}

The measured and expected \ZZ and \ZH signal strengths are reported in terms of the signal strength modifier $\mu$, defined as the ratio of the value of the cross section to the expected SM theoretical cross section, $\sigma/\sigma_{\text{SM}}$.
The \CLs method~\cite{CLS2,CLS1} is used to determine the upper limits on the signal strengths at 95\% confidence level (CL).

The final fit procedure is validated using synthetic data samples without statistical fluctuations, and also by treating  one of the mixed models as the observed four-tag data.
In both tests, the behavior of the systematic uncertainties is as expected and the resulting best fit signal strengths were consistent with zero.

\begin{figure}[hbt!]\centering
  \includegraphics[width=0.49\textwidth]{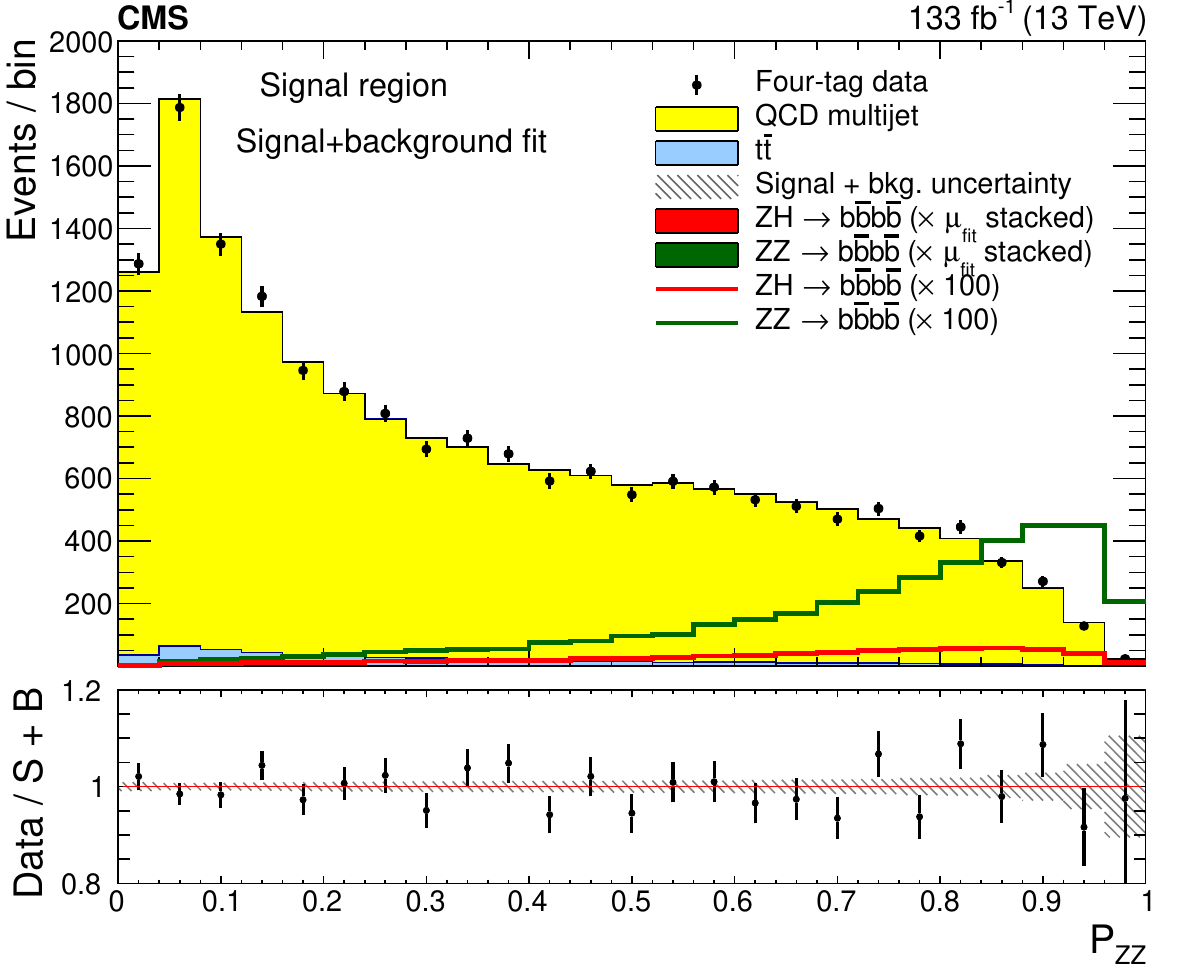}\hfill
  \includegraphics[width=0.49\textwidth]{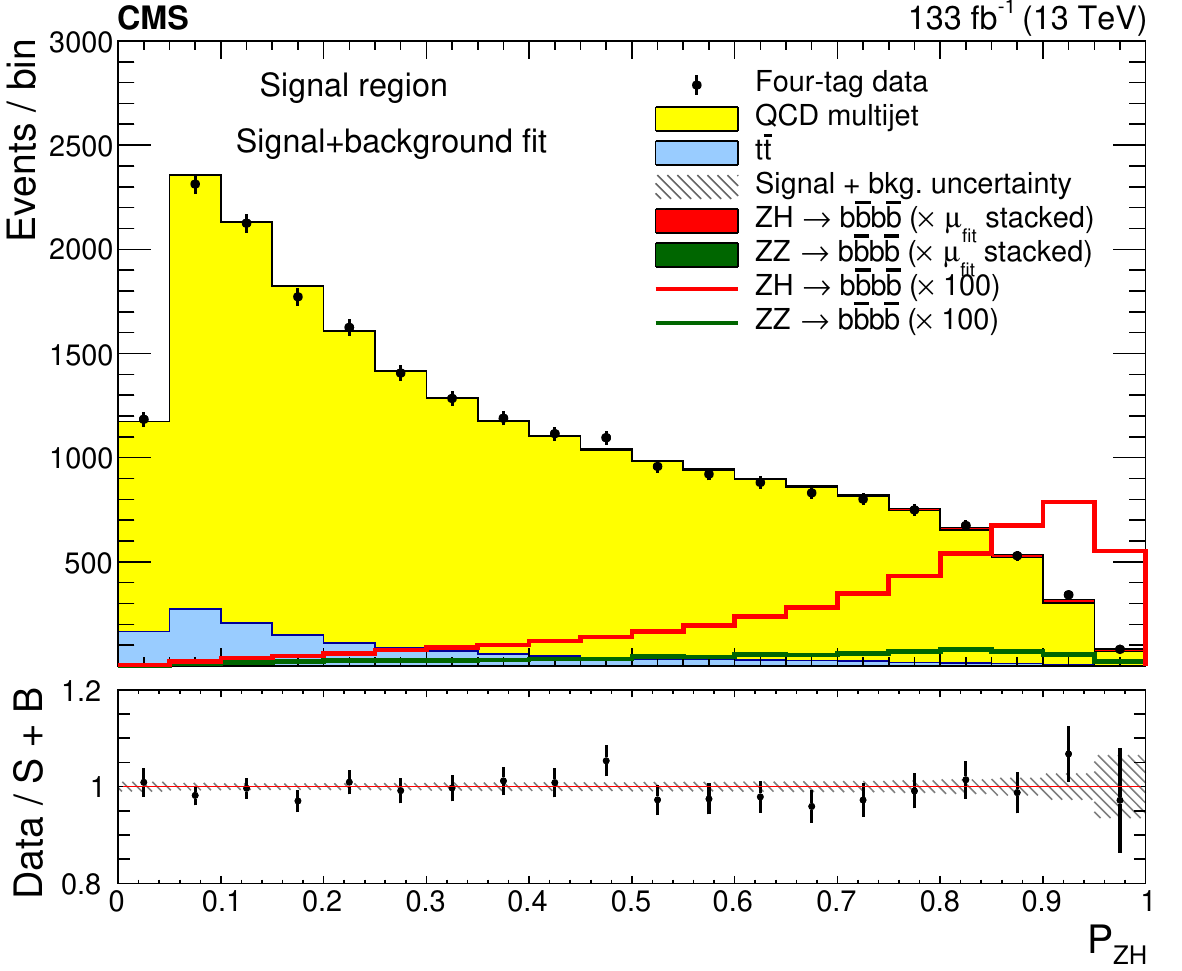}  
  \caption{
  Distributions of signal probabilities for \ZZ (\cmsLeft) and \ZH (\cmsRight) channels (points), along with the post-fit QCD multijet (yellow region) plus \ttbar\ (blue region) distributions.
  The \ZH and \ZZ signal distributions scaled to the fitted signal strengths are shown, stacked on top of the background prediction.
  The expected \ZH (red histograms) and \ZZ (green histograms) signal channel distributions are also shown separately, multiplied by 100 for visibility.
  The lower panels display the ratio of the data to the result of the signal plus background fit, with the hatched area showing the uncertainty in the combined fit.
  }
  \label{fig:postfit_ZZ_ZH}
\end{figure}

Figure~\ref{fig:postfit_ZZ_ZH} shows the results of the combined fit of the SvB signal probability distribution to the signal plus background model.
The resulting signal strengths and corresponding 95\% CL upper limits are shown in Table~\ref{tab:expected_SvB_MA}.
The values if only the statistical uncertainties are included are shown in parentheses.
The upper limits are determined under the assumption that no signal exists.
Despite the significant difference in the \ZZ and \ZH cross sections, the upper limits of the signal strengths are similar.
This is due to the increased signal selection efficiency and lower background levels in the SR for the \ZH channel in comparison to those for the \ZZ channel.
The current results are limited by the size of the data set and the systematic uncertainties associated with the background model.

\begin{table*}[htb!]
\topcaption{Expected and observed \ZZ and \ZH signal strengths and their corresponding 95\% CL upper limits.
The expected signal strengths and the corresponding expected upper limits shown in parentheses include only the statistical uncertainties. 
The upper limits are obtained from a fit to the SvB signal probabilities under the hypothesis of no $\ZZ\to 4\PQb$ or $\ZH\to 4\PQb$ signal.
}\label{tab:expected_SvB_MA}
\renewcommand{\arraystretch}{1.2}
\centering
\begin{tabular}{lcc}
                     & \ZZ  & \ZH    \\
\hline\\[-2.2ex]
Signal strength expected (stat. only) & $1.0\ _{-1.7}^{+1.9}$ ($1.0\ _{-1.3}^{+1.4}$) & $1.0\ _{-1.4}^{+1.5}$ ($1.0\ _{-1.1}^{+1.1}$)  \\
Signal strength observed  & $0.0\ _{-1.7}^{+2.0}$  & $2.2\ _{-0.8}^{+0.9}$    \\
Expected upper limit at $95\%$ CL (stat. only) & 3.8 (2.8) & 2.9 (2.3)   \\
Observed upper limit at $95\%$ CL & 3.8  & 5.0   \\
\end{tabular}
\end{table*}

\section{Summary}\label{sec:summary}

A search for \ZZ and \ZH production in the 4\PQb final state is presented.
The search uses the full 2016--2018 data set of proton-proton collisions at a center-of-mass energy of 13\TeV recorded with the CMS detector at the LHC, corresponding to an integrated luminosity of 133\fbinv.
The analysis benefits from a multiclass multivariate classifier, which uses convolutions to solve the combinatoric jet pairing problem, and has been designed with an architecture customized to the 4\PQb final state.
The classifier is used both for signal-versus-background discrimination and for the derivation and validation of the background model.
A novel technique for assessing the background modeling uncertainties, using a synthetic data sample, produced using a hemisphere mixing procedure, allows both the uncertainty in the background model and its variance to be measured with a precision better than the statistical uncertainties in the selected signal-region events.
While these techniques are developed and demonstrated in the \ZZ and \ZH $\to$ 4\PQb searches, they are directly applicable to the $\HH \to 4\PQb$ analysis.
The observed (expected) 95\% CL upper limits on the $\ZZ \to 4\PQb$ and $\ZH \to 4\PQb$ production cross sections correspond to 3.8 (3.8) and 5.0 (2.9) times the standard model prediction, respectively.

\begin{acknowledgments}
  We congratulate our colleagues in the CERN accelerator departments for the excellent performance of the LHC and thank the technical and administrative staffs at CERN and at other CMS institutes for their contributions to the success of the CMS effort. In addition, we gratefully acknowledge the computing centers and personnel of the Worldwide LHC Computing Grid and other centers for delivering so effectively the computing infrastructure essential to our analyses. Finally, we acknowledge the enduring support for the construction and operation of the LHC, the CMS detector, and the supporting computing infrastructure provided by the following funding agencies: SC (Armenia), BMBWF and FWF (Austria); FNRS and FWO (Belgium); CNPq, CAPES, FAPERJ, FAPERGS, and FAPESP (Brazil); MES and BNSF (Bulgaria); CERN; CAS, MoST, and NSFC (China); MINCIENCIAS (Colombia); MSES and CSF (Croatia); RIF (Cyprus); SENESCYT (Ecuador); ERC PRG, RVTT3 and MoER TK202 (Estonia); Academy of Finland, MEC, and HIP (Finland); CEA and CNRS/IN2P3 (France); SRNSF (Georgia); BMBF, DFG, and HGF (Germany); GSRI (Greece); NKFIH (Hungary); DAE and DST (India); IPM (Iran); SFI (Ireland); INFN (Italy); MSIP and NRF (Republic of Korea); MES (Latvia); LMTLT (Lithuania); MOE and UM (Malaysia); BUAP, CINVESTAV, CONACYT, LNS, SEP, and UASLP-FAI (Mexico); MOS (Montenegro); MBIE (New Zealand); PAEC (Pakistan); MES and NSC (Poland); FCT (Portugal);  MESTD (Serbia); MCIN/AEI and PCTI (Spain); MOSTR (Sri Lanka); Swiss Funding Agencies (Switzerland); MST (Taipei); MHESI and NSTDA (Thailand); TUBITAK and TENMAK (Turkey); NASU (Ukraine); STFC (United Kingdom); DOE and NSF (USA).

\hyphenation{Rachada-pisek} Individuals have received support from the Marie-Curie program and the European Research Council and Horizon 2020 Grant, contract Nos.\ 675440, 724704, 752730, 758316, 765710, 824093, 101115353,101002207, and COST Action CA16108 (European Union); the Leventis Foundation; the Alfred P.\ Sloan Foundation; the Alexander von Humboldt Foundation; the Science Committee, project no. 22rl-037 (Armenia); the Belgian Federal Science Policy Office; the Fonds pour la Formation \`a la Recherche dans l'Industrie et dans l'Agriculture (FRIA-Belgium); the Agentschap voor Innovatie door Wetenschap en Technologie (IWT-Belgium); the F.R.S.-FNRS and FWO (Belgium) under the ``Excellence of Science -- EOS" -- be.h project n.\ 30820817; the Beijing Municipal Science \& Technology Commission, No. Z191100007219010 and Fundamental Research Funds for the Central Universities (China); the Ministry of Education, Youth and Sports (MEYS) of the Czech Republic; the Shota Rustaveli National Science Foundation, grant FR-22-985 (Georgia); the Deutsche Forschungsgemeinschaft (DFG), under Germany's Excellence Strategy -- EXC 2121 ``Quantum Universe" -- 390833306, and under project number 400140256 - GRK2497; the Hellenic Foundation for Research and Innovation (HFRI), Project Number 2288 (Greece); the Hungarian Academy of Sciences, the New National Excellence Program - \'UNKP, the NKFIH research grants K 131991, K 133046, K 138136, K 143460, K 143477, K 146913, K 146914, K 147048, 2020-2.2.1-ED-2021-00181, and TKP2021-NKTA-64 (Hungary); the Council of Science and Industrial Research, India; ICSC -- National Research Center for High Performance Computing, Big Data and Quantum Computing, funded by the NextGenerationEU program (Italy); the Latvian Council of Science; the Ministry of Education and Science, project no. 2022/WK/14, and the National Science Center, contracts Opus 2021/41/B/ST2/01369 and 2021/43/B/ST2/01552 (Poland); the Funda\c{c}\~ao para a Ci\^encia e a Tecnologia, grant CEECIND/01334/2018 (Portugal); the National Priorities Research Program by Qatar National Research Fund; MCIN/AEI/10.13039/501100011033, ERDF ``a way of making Europe", and the Programa Estatal de Fomento de la Investigaci{\'o}n Cient{\'i}fica y T{\'e}cnica de Excelencia Mar\'{\i}a de Maeztu, grant MDM-2017-0765 and Programa Severo Ochoa del Principado de Asturias (Spain); the Chulalongkorn Academic into Its 2nd Century Project Advancement Project, and the National Science, Research and Innovation Fund via the Program Management Unit for Human Resources \& Institutional Development, Research and Innovation, grant B37G660013 (Thailand); the Kavli Foundation; the Nvidia Corporation; the SuperMicro Corporation; the Welch Foundation, contract C-1845; and the Weston Havens Foundation (USA).
\end{acknowledgments}

\bibliography{auto_generated} 
\cleardoublepage \appendix\section{The CMS Collaboration \label{app:collab}}\begin{sloppypar}\hyphenpenalty=5000\widowpenalty=500\clubpenalty=5000
\cmsinstitute{Yerevan Physics Institute, Yerevan, Armenia}
{\tolerance=6000
A.~Hayrapetyan, A.~Tumasyan\cmsAuthorMark{1}\cmsorcid{0009-0000-0684-6742}
\par}
\cmsinstitute{Institut f\"{u}r Hochenergiephysik, Vienna, Austria}
{\tolerance=6000
W.~Adam\cmsorcid{0000-0001-9099-4341}, J.W.~Andrejkovic, T.~Bergauer\cmsorcid{0000-0002-5786-0293}, S.~Chatterjee\cmsorcid{0000-0003-2660-0349}, K.~Damanakis\cmsorcid{0000-0001-5389-2872}, M.~Dragicevic\cmsorcid{0000-0003-1967-6783}, P.S.~Hussain\cmsorcid{0000-0002-4825-5278}, M.~Jeitler\cmsAuthorMark{2}\cmsorcid{0000-0002-5141-9560}, N.~Krammer\cmsorcid{0000-0002-0548-0985}, A.~Li\cmsorcid{0000-0002-4547-116X}, D.~Liko\cmsorcid{0000-0002-3380-473X}, I.~Mikulec\cmsorcid{0000-0003-0385-2746}, J.~Schieck\cmsAuthorMark{2}\cmsorcid{0000-0002-1058-8093}, R.~Sch\"{o}fbeck\cmsorcid{0000-0002-2332-8784}, D.~Schwarz\cmsorcid{0000-0002-3821-7331}, M.~Sonawane\cmsorcid{0000-0003-0510-7010}, S.~Templ\cmsorcid{0000-0003-3137-5692}, W.~Waltenberger\cmsorcid{0000-0002-6215-7228}, C.-E.~Wulz\cmsAuthorMark{2}\cmsorcid{0000-0001-9226-5812}
% ...
\par}
\vskip\cmsinstskip
\dag:~Deceased\\
$^{1}$Also at Yerevan State University, Yerevan, Armenia\\
$^{2}$Also at TU Wien, Vienna, Austria\\
% ...
\end{sloppypar}
\end{document}